\documentclass[12pt, onecolumn, draftclsnofoot]{IEEEtran}%

\usepackage[T1]{fontenc}
\usepackage{amsmath,amssymb,amsfonts,mathrsfs,bm}
\usepackage{mathtools}
\usepackage{amsthm}
\usepackage{nicefrac}
\usepackage{cite}
\usepackage[shortlabels]{enumitem}
\usepackage{graphicx}
\usepackage{epstopdf}
\usepackage{url}
\usepackage{colortbl}
\usepackage{booktabs}
\usepackage{multirow}
\usepackage[table,dvipsnames]{xcolor}
\usepackage[normalem]{ulem}

\usepackage{array}
\newcolumntype{L}[1]{>{\raggedright\let\newline\\\arraybackslash\hspace{0pt}}m{#1}}
\newcolumntype{C}[1]{>{\centering\let\newline\\\arraybackslash\hspace{0pt}}m{#1}}
\newcolumntype{R}[1]{>{\raggedleft\let\newline\\\arraybackslash\hspace{0pt}}m{#1}}

\makeatletter
\let\MYcaption\@makecaption
\makeatother
\usepackage[font=footnotesize]{subcaption}
\makeatletter
\let\@makecaption\MYcaption
\makeatother

\usepackage{xparse}



\usepackage{glossaries}

\makeatletter
\let\oldgls\gls
\let\oldglspl\glspl

\newcommand\fussy@ifnextchar[3]{%
  \let\reserved@d=#1%
  \def\reserved@a{#2}%
  \def\reserved@b{#3}%
  \futurelet\@let@token\fussy@ifnch}
\def\fussy@ifnch{%
  \ifx\@let@token\reserved@d
    \let\reserved@c\reserved@a 
  \else
    \let\reserved@c\reserved@b
  \fi
 \reserved@c}

\renewcommand{\gls}[1]{%
  \oldgls{#1}\fussy@ifnextchar.{\@checkperiod}{\@}}
\renewcommand{\glspl}[1]{%
  \oldglspl{#1}\fussy@ifnextchar.{\@checkperiod}{\@}}

\newcommand{\@checkperiod}[1]{%
  \ifnum\sfcode`\.=\spacefactor\else#1\fi
}
\makeatother

\newacronym{wrt}{w.r.t.}{with respect to}
\newacronym{RHS}{RHS}{right-hand side}
\newacronym{LHS}{LHS}{left-hand side}
\newacronym{iid}{i.i.d.}{independent and identically distributed}

\usepackage{float}

\ifx\notloadhyperref\undefined
	\ifx\loadbibentry\undefined
		\usepackage[hidelinks,hypertexnames=false]{hyperref} 
	\else
		\usepackage{bibentry}
		\makeatletter\let\saved@bibitem\@bibitem\makeatother
		\usepackage[hidelinks,hypertexnames=false]{hyperref}
		\makeatletter\let\@bibitem\saved@bibitem\makeatother
	\fi
\else
	\ifx\loadbibentry\undefined
		\relax
	\else
		\usepackage{bibentry}
	\fi
\fi

\usepackage[capitalize]{cleveref}
\crefname{equation}{}{}
\Crefname{equation}{}{}
\crefname{claim}{claim}{claims}
\crefname{step}{step}{steps}
\crefname{line}{line}{lines}
\crefname{condition}{condition}{conditions}
\crefname{dmath}{}{}
\crefname{dseries}{}{}
\crefname{dgroup}{}{}

\crefname{Problem}{Problem}{Problems}
\crefformat{Problem}{Problem~(#2#1#3)}
\crefrangeformat{Problem}{Problems~(#3#1#4) to~(#5#2#6)}

\crefname{Theorem}{Theorem}{Theorems}
\crefname{Corollary}{Corollary}{Corollaries}
\crefname{Proposition}{Proposition}{Propositions}
\crefname{Lemma}{Lemma}{Lemmas}
\crefname{Definition}{Definition}{Definitions}
\crefname{Example}{Example}{Examples}
\crefname{Assumption}{Assumption}{Assumptions}
\crefname{Remark}{Remark}{Remarks}
\crefname{Rem}{Remark}{Remarks}
\crefname{remarks}{Remarks}{Remarks}
\crefname{Appendix}{Appendix}{Appendices}
\crefname{Supplement}{Supplement}{Supplements}
\crefname{Exercise}{Exercise}{Exercises}
\crefname{Theorem_A}{Theorem}{Theorems}
\crefname{Corollary_A}{Corollary}{Corollaries}
\crefname{Proposition_A}{Proposition}{Propositions}
\crefname{Lemma_A}{Lemma}{Lemmas}
\crefname{Definition_A}{Definition}{Definitions}

\usepackage{crossreftools}
\ifx\notloadhyperref\undefined
\else
	\relax
\fi

\usepackage{algorithm,algorithmic}

\ifx\loadbreqn\undefined
	\relax
\else
	\usepackage{breqn} 
\fi

\ifx\renewtheorem\undefined
\ifx\useTheoremCounter\undefined
\newtheorem{Theorem}{Theorem}
\newtheorem{Corollary}{Corollary}
\newtheorem{Proposition}{Proposition}
\newtheorem{Lemma}{Lemma}
\else
\newtheorem{Theorem}{Theorem}

\newtheorem{Proposition}[theorem]{Proposition}
\fi

\newtheorem{Definition}{Definition}
\newtheorem{Example}{Example}

\fi

\theoremstyle{remark}

\theoremstyle{plain}

\newcommand{\qednew}{\nobreak \ifvmode \relax \else
      \ifdim\lastskip<1.5em \hskip-\lastskip
      \hskip1.5em plus0em minus0.5em \fi \nobreak
      \vrule height0.75em width0.5em depth0.25em\fi}

\newcommand{\Real}{\mathbb{R}}

\newcommand{\calE}{\mathcal{E}}

\newcommand{\calI}{\mathcal{I}}

\newcommand{\calO}{\mathcal{O}}

\newcommand{\scF}{\mathscr{F}}

\DeclareSymbolFont{bsfletters}{OT1}{cmss}{bx}{n}
\DeclareSymbolFont{ssfletters}{OT1}{cmss}{m}{n}
\DeclareMathSymbol{\bsfGamma}{0}{bsfletters}{'000}
\DeclareMathSymbol{\ssfGamma}{0}{ssfletters}{'000}
\DeclareMathSymbol{\bsfDelta}{0}{bsfletters}{'001}
\DeclareMathSymbol{\ssfDelta}{0}{ssfletters}{'001}
\DeclareMathSymbol{\bsfTheta}{0}{bsfletters}{'002}
\DeclareMathSymbol{\ssfTheta}{0}{ssfletters}{'002}
\DeclareMathSymbol{\bsfLambda}{0}{bsfletters}{'003}
\DeclareMathSymbol{\ssfLambda}{0}{ssfletters}{'003}
\DeclareMathSymbol{\bsfXi}{0}{bsfletters}{'004}
\DeclareMathSymbol{\ssfXi}{0}{ssfletters}{'004}
\DeclareMathSymbol{\bsfPi}{0}{bsfletters}{'005}
\DeclareMathSymbol{\ssfPi}{0}{ssfletters}{'005}
\DeclareMathSymbol{\bsfSigma}{0}{bsfletters}{'006}
\DeclareMathSymbol{\ssfSigma}{0}{ssfletters}{'006}
\DeclareMathSymbol{\bsfUpsilon}{0}{bsfletters}{'007}
\DeclareMathSymbol{\ssfUpsilon}{0}{ssfletters}{'007}
\DeclareMathSymbol{\bsfPhi}{0}{bsfletters}{'010}
\DeclareMathSymbol{\ssfPhi}{0}{ssfletters}{'010}
\DeclareMathSymbol{\bsfPsi}{0}{bsfletters}{'011}
\DeclareMathSymbol{\ssfPsi}{0}{ssfletters}{'011}
\DeclareMathSymbol{\bsfOmega}{0}{bsfletters}{'012}
\DeclareMathSymbol{\ssfOmega}{0}{ssfletters}{'012}

\DeclareMathOperator*{\argmin}{arg\,min}

\DeclareMathOperator{\st}{s.t.}

\DeclareMathOperator{\Mod}{mod}

\DeclarePairedDelimiter\abs{\lvert}{\rvert}
\DeclarePairedDelimiter\parens{(}{)}

\DeclarePairedDelimiter\braces{\{}{\}}

\DeclarePairedDelimiterX\ip[2]{\langle}{\rangle}{#1,#2}
\DeclarePairedDelimiterX\norm[1]{\lVert}{\rVert}{#1}
\DeclarePairedDelimiterXPP\col[1]{\operatorname{col}}{\{}{\}}{}{#1} 
\DeclarePairedDelimiterXPP\row[1]{\operatorname{row}}{\{}{\}}{}{#1} 
\DeclarePairedDelimiterXPP\erf[1]{\operatorname{erf}}{(}{)}{}{#1}
\DeclarePairedDelimiterXPP\erfc[1]{\operatorname{erfc}}{(}{)}{}{#1}
\DeclarePairedDelimiterXPP\op[2]{\operatorname{#1}}{(}{)}{}{#2} 

\newcommand{\T}{^{\intercal}}
\newcommand{\ud}{\,\mathrm{d}} 

\newcommand{\ofrac}[1]{{\frac{1}{#1}}}

\newcommand{\floor}[1]{\left\lfloor{#1}\right\rfloor}

\DeclarePairedDelimiterX\Set[2]\{\}{%

#2
}
\DeclarePairedDelimiterX\Setc[1]\{\}{%

#1
}

\NewDocumentCommand\set{s o m}{%
	\IfBooleanTF#1%
	{\IfValueTF{#2}{\Set*{#2}{#3}}{\Setc*{#3}}}%
	{\IfValueTF{#2}{\Set{#2}{#3}}{\Setc{#3}}}%
}

\NewDocumentCommand{\evalat}{s O{\big} m m}{%
  \IfBooleanTF{#1}
   {\left. #3 \right|_{#4}}
   {#3#2|_{#4}}%
}

\NewDocumentCommand \ifcond {m m} {%
	{#1} %
	\IfValueT{#2}{\, \middle|\, {#2}}%
}

\DeclareDocumentCommand \P {e{_} g >{\SplitArgument{ 1 }{ @| }}d() g } {%
	\mathbb{P}%
	\IfValueTF{#1}{_{#1}}
		{\IfValueT{#2}{_{#2}}}%
	\IfValueT{#3}{\left(\ifcond#3}%
	\IfValueT{#4}{\, \middle|\, {#4}}%
	\IfValueT{#3}{\right)}%
}

\DeclareDocumentCommand \E {e{_} g >{\SplitArgument{ 1 }{ @| }}o g } {%
	\mathbb{E}%
	\IfValueTF{#1}{_{#1}}
		{\IfValueT{#2}{_{#2}}}%
	\IfValueT{#3}{\left[\ifcond#3}%
	\IfValueT{#4}{\, \middle|\, {#4}}%
	\IfValueT{#3}{\right]}%
}

\renewcommand{\figurename}{Fig.}
\newcommand{\figref}[1]{\figurename~\ref{#1}}
\graphicspath{{./Figures/}} 
\newcommand{\includeCroppedPdf}[2][]{%
    \IfFileExists{./Figures/#2-crop.pdf}{}{%
        \immediate\write18{pdfcrop ./Figures/#2 ./Figures/#2-crop.pdf}}%
    \includegraphics[#1]{./Figures/#2-crop.pdf}}

\newcommand{\beginsupplement}{
    \setcounter{section}{0}
    \renewcommand{\thesection}{S\arabic{section}}
    \setcounter{equation}{0}
    \renewcommand{\theequation}{S\arabic{equation}}
    \setcounter{table}{0}
    \renewcommand{\thetable}{S\arabic{table}}
    \setcounter{figure}{0}
    \renewcommand{\thefigure}{S\arabic{figure}}
}

\definecolor{gray90}{gray}{0.9}

\ifx\nohighlights\undefined

	\newcommand{\msout}[1]{\text{\color{green} \sout{\ensuremath{#1}}}}
	\newcommand{\del}[1]{{\color{green}\ifmmode \msout{#1}\else\sout{#1}\fi}}
\else

	\newcommand{\msout}[1]{#1}
	\newcommand{\del}[1]{#1}
\fi

\newcommand{\hhide}[1]{}

\ifx\diagnoselabel\undefined
	\relax
\else
	\makeatletter
	 \def\@testdef #1#2#3{%
		 \def\reserved@a{#3}\expandafter \ifx \csname #1@#2\endcsname
		\reserved@a  \else
	 \typeout{^^Jlabel #2 changed:^^J%
	 \meaning\reserved@a^^J%
	 \expandafter\meaning\csname #1@#2\endcsname^^J}%
	 \@tempswatrue \fi}
	\makeatother
\fi

\usepackage{tikz}
\usepackage{pgfplots}
\pgfplotsset{compat=1.5}

\providecommand{\U}[1]{\protect\rule{.1in}{.1in}}

\def\ui{{\mathrm{i}}}

\newcommand{\PV}[1]{P_{V_{#1}}}
\newcommand{\oPV}[1]{\overline{P}_{V_{#1}}}
\newcommand{\FV}[1]{\scF_{V_{#1}}}

\theoremstyle{definition}

\newtheorem{prob}{Problem}

\crefname{prob}{Problem}{Problems}

\begin{document}
\date{}
\title{Subgraph Signal Processing}
\author{Feng~Ji, Wee~Peng~Tay,~\IEEEmembership{Senior Member,~IEEE} and Giacomo~Kahn}%
\maketitle
\begin{abstract}
Graph signal processing, like the graph Fourier transform, requires the full graph signal at every vertex of the graph. However, in practice, only signals at a subset of vertices may be available. We propose a subgraph signal processing framework that relates a graph shift operator or filter on a subgraph with a filter on the ambient graph through an operator loss. It allows us to define shift operators for the subgraph signal, which has a meaningful interpretation and relation to mixtures of shift invariant filters restricted to different subgraphs of the ambient graph (which we call semi shift invariant). This leads to a notion of frequency domain for the subgraph signal consistent in some sense with that of the full graph signal. We apply the subgraph signal processing machinery to several applications and demonstrate the utility of this framework in cases where only partial graph signals are observed.
\end{abstract}


\section{Introduction}

Since its emergence, the theory and applications of graph signal processing (GSP) have rapidly developed. GSP incorporates geometric properties of a graph in analyzing signals supported on it. The theory covers a wide range of topics with a full array of applications, including graph signal filtering, downsampling, clustering, detection, graph time series and learning with graph neural networks \cite{Shu13, San13, San14, Gad14, Don16, Def16,Kip16, Egi17, Sha17, Gra18, Ort18, Girault2018, JiTay:J19, Suc17, Wan18, JiYanZha:C20}.

The heart and soul of the theory is the notion of a \emph{ graph shift operator}. Once an appropriate graph shift operator is chosen, standard signal processing tasks like Fourier transform, filtering and sampling can be defined (see for example \cite{Shu13,JiTay:J19} for more details). To capture the correlation between signals on vertices of the graph, such a shift operator should be associated with the topology of the graph, e.g., imitating a diffusion process on the graph. In addition, it should have a few prescribed algebraic properties, e.g., admitting a real eigenbasis (a basis consisting of eigenvectors). Common examples include the adjacency matrix and graph Laplacian. The decomposition of a signal \gls{wrt} an eigenbasis of the graph shift operator gives rise to the \emph{graph Fourier transform (GFT)}, and the coordinates \gls{wrt} this basis is the \emph{frequency domain}, analogous to its counterpart in the classical theory of discrete Fourier transform (DFT). As a consequence, we are able to give a different interpretation of a graph signal $x$ by looking at its corresponding transform $\hat{x}$ in the frequency domain.

In order to compute the frequency spectrum $\hat{x}$, one needs to make use of every entry of $x$, i.e., the signal value at every vertex of the graph. However, there can be scenarios where full observation of $x$ is impossible. For example, in a sensor network, it is possible that readings from some sensors are missing due to reasons such as processing delay \cite{Kha16}, damage, energy conservation \cite{TayTsiWin:J07a}, or lack of access due to reasons such as privacy \cite{SunTay:J20a,SunTay:J20b,WanSonTay:J21}. As another example, if we view an image as a two dimensional grid graph and pixel values as signals, some of the pixels can be missing due to corruption of the image. As a partial observation of $x$ is supported on a subset of vertices of the graph, it is called a \emph{subgraph signal}. In this paper, we propose a signal processing framework that allows one to make meaningful interpretation of subgraph signals, by finding a frequency domain ``consistent'' with the original underlying graph signal. 

\begin{figure}[!htb]
\centering
\includegraphics[width=1\linewidth]{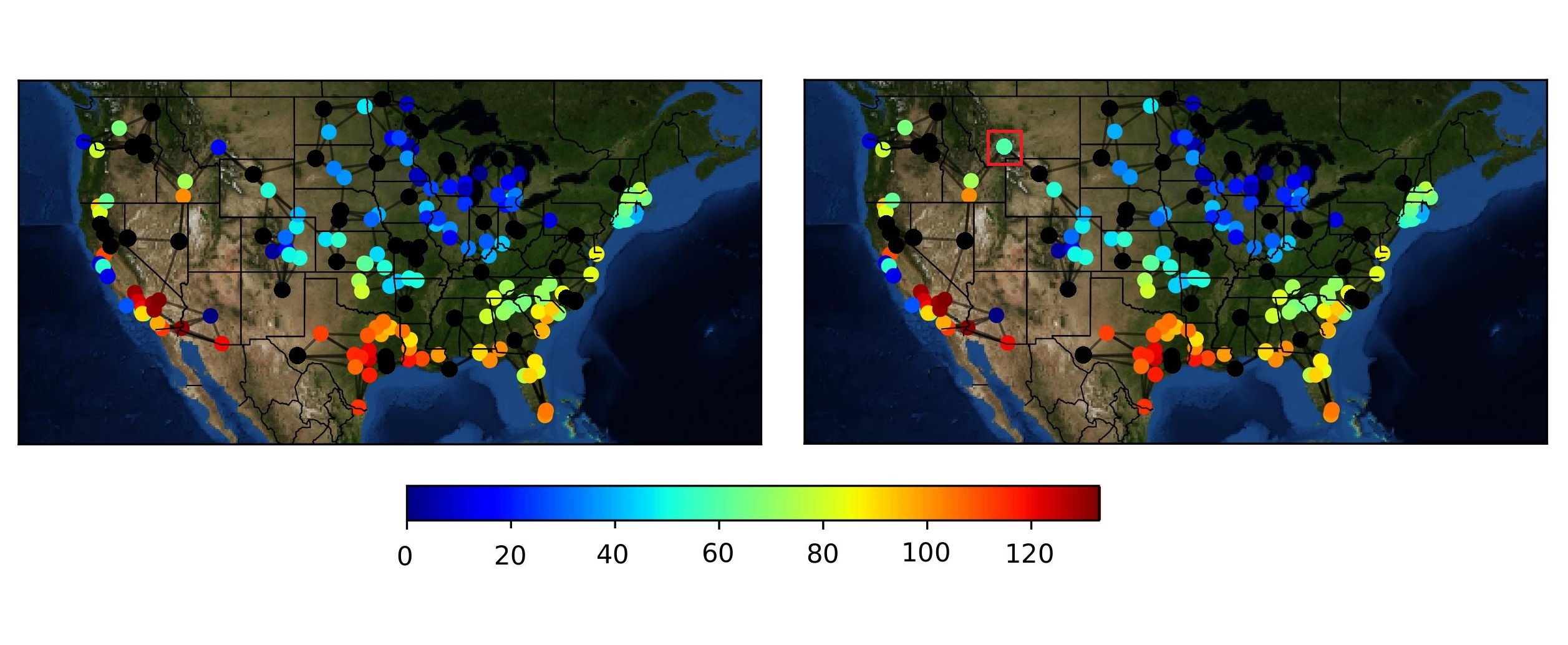}
\caption{Actual (left) and perturbed (right) temperature signals in a network of weather stations in the United States.}
\label{fig:ssp13}
\end{figure}

\begin{Example} \label{eg:fsa}
\cref{fig:ssp13} shows a network of weather stations in the United States. The colored nodes in the left figure indicate the actual daily temperature readings. However, the readings for a large subset of stations are missing (as shown by the black nodes). With only partially observed information, we wish to perform signal processing tasks that allow us to perform inference similar to the scenario where we have full observations. For example, suppose there is an abnormally high temperature at a single station $v$, where $v$ is highlighted by the red square in the right figure. Using only the available readings from stations in the close vicinity, the temperature at $v$ does not seem abnormal. Therefore, in order to detect the anomaly, it might be insufficient to consider the partial observations we have from the one-hop neighborhood of $v$, which contains only two other nodes in the subgraph with observed temperature readings. Instead, we need to consider more stations in the northern part of the country, which suggests that the reading should be lower (as shown on the left). Our goal is to develop a subgraph signal processing framework to allow inferences on the partial observations that are ``consistent'' with the underlying graph signal. Further discussion and more details for this particular example are provided in \cref{sec:sim}. 
\end{Example}

\begin{figure}[!htb]
\centering
\includegraphics[width=0.65\textwidth]{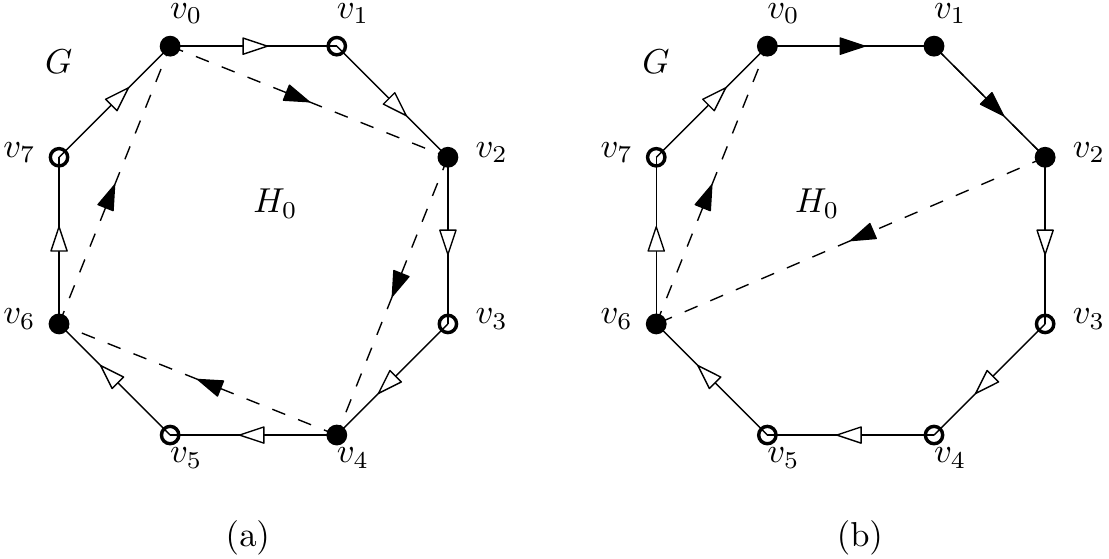}
\caption{The subset $V_0$ consists of vertices indicated by the disks. The edges of a subgraph $H_0$ on $V_0$ are indicated by the dashed lines.}\label{fig:subgraph}
\end{figure}

\begin{Example}\label{eg:subgraph}
Consider the subset of vertices $V_0$ of the unweighted directed graph $G$ in \cref{fig:subgraph}a. Suppose that we consider the adjacency matrix $A_G$ as the graph shift operator for $G$. In \cref{fig:subgraph}a, the adjacency matrix $A_{H_0}$ of the subgraph $H_0$ shown is a submatrix consisting of the even-indexed rows and columns of $A_G^2$. Any filter (i.e., a linear transformation on the graph signal) shift invariant \gls{wrt} $A_{H_0}$ can be viewed as a polynomial of $A_G^2$ composed with a projection to the vertices of $H_0$. However, in \cref{fig:subgraph}b, the adjacency matrix $A_{H_0}$ of the subgraph $H_0$ cannot be expressed as the projection of a polynomial of $A$. A natural question that arises is what filter on $G$ mimics the effect of $A_{H_0}$ up to a certain ``consistency''? This filter is necessarily not shift invariant \gls{wrt} $A_G$. Removing the restriction to adjacency matrices, we may ask what shift operators or shift invariant filters for $H_0$ are in some sense ``consistent'' with filters on $G$? Insights to these questions allow us to translate the effects of a shift invariant filter learned from signals on $V_0$ to the bigger ambient graph $G$, and vice versa. 
\end{Example}

Our main contributions in this paper are the following:
\begin{enumerate}
	\item We formulate the concept of a subgraph signal processing (subGSP) transformation pair that relates a filter on a subgraph with another filter on the given ambient graph. We explicitly define the notion of ``consistency'' allured to in the above examples through the introduction of filter families and an operator loss.
	\item We introduce the class of semi shift invariant filters, which are essentially mixtures of shift invariant filters restricted to different subgraphs of the ambient graph. We derive several properties of this filter family, which indicate that it is a sufficiently rich family for our subGSP framework. 
	\item We propose suitable choices for the filter families in our subGSP framework and formulate several problems that one commonly encounters in practice, using the machinery of the subGSP framework. We demonstrate how these formulations can be applied in practice through numerical examples. Numerical results suggest that the subGSP approach yields better performance than other graph reduction techniques like Kron reduction.  
\end{enumerate}
A preliminary version outlining our framework was submitted to the 46th International Conference on Acoustics, Speech, \& Signal Processing. In this paper, we provide technical details, proofs of technical results, examples and discussions and more applications of the subGSP framework.

A common graph reduction technique is the Kron reduction \cite{Dor13}, which was originally developed in circuit theory to reduce large circuits to electrically equivalent smaller circuits. The Kron reduced graph \gls{wrt} to a subset of vertices $V_0$ enjoys various topological and spectral properties \cite{Dor13}. However, a filter shift invariant \gls{wrt} the Kron reduced Laplacian is in general not related or approximated well (after taking a projection) by shift invariant filters (or a mixture of them) under the ambient graph Laplacian. This makes it difficult to interpret the consistency of the Kron reduced Laplacian \gls{wrt} the ambient graph Laplacian. Various signal processing tasks also achieve better performance in our proposed framework because the Kron reduced Laplacian belongs to the family of operators that our framework optimizes over when identifying a suitable subGSP transformation pair. We provide more detailed discussions and comparisons with our proposed framework in \cref{sec:sga,sec:sim}.

Our objective might seem similar to graph sampling \cite{Aga13, Che15, Tsi15, Mar16, Anis2016}. However, graph sampling assumes prior knowledge of properties of the full graph signal $x$ such as band-limitedness. With such knowledge, one selects an optimal subset of vertices such that the restriction of $x$ at these sample vertices captures most of or even full information of $x$. On the other hand, in this paper, the subset of vertices is fixed \emph{a priori} regardless of the signal $x$. The restriction of $x$ to the given subset may not carry full information of $x$, which prevents us from using GSP on the entire graph. Just like GSP, the theory revolves around a chosen operator. In GSP, key theoretical concepts such as Fourier transform and frequency domain are defined without referring to specific signals, and the only input is the graph topology. The subGSP framework is in the same spirit by only making use of the geometry, i.e., the embedding of the subgraph.

Another related problem is sparse signal interpolation \cite{Don03, Can06, Zhu12}, in which a sparse prior of the graph signal is assumed. These approaches are based on tools such as compressed sensing. In this paper, we want to develop a set of generic signal processing tools not targeting or referring to any specific family of signals, just as GSP, which is a generic framework for any graph signals. Apart from the graph structure, a user of subGSP tools does not have to know anything of the ambient graph signal.

The rest of this paper is organized as follows. In \cref{sec:sub}, we formally introduce the problem of subGSP on a given subset of vertices in a graph. As we recalled briefly above, the essential step is to find an operator $F$ as a shift on the given subset. We summarize some of the signal processing tasks one can perform once such an operator is found. To obtain $F$, we need to search within a family of operators, called \emph{semi shift invariant filters}, as a generalization of several familiar filter families. We introduce these in \cref{sec:semi} and discuss their main theoretical properties. In \cref{sec:sga}, we discuss how to cast the problem of finding $F$ as an optimization problem. We present simulation results in \cref{sec:sim} and conclude in \cref{sec:con}.  

\section{What is subgraph signal processing?} \label{sec:sub}

In this section, we describe a general subGSP framework cum problem formulation. Let $G = (V,E)$ denote a graph, where $V$ is the set of vertices and $E$ the set of (weighted) edges. Graph signals on the graph $G$ can be canonically identified with $\mathbb{R}^{|V|}$, denoted by $L^2(V)$ with norm $\norm{\cdot}_2$. A \emph{filter} on $L^2(V)$ is a linear linear map $L^2(V)\mapsto L^2(V)$. As all filters on $L^2(V)$ can be represented by matrices given fixed bases, we abuse terminology by using these terms interchangeably. An eigenbasis of $L^2(V)$ is a basis consisting of eigenvectors of a given operator.

Let $V_0 \subset V$. The signals on $V_0$ are identified with $\mathbb{R}^{|V_0|}$, denoted by $L^2(V_0)$. There is an obvious restriction map $P_{V_0}: L^2(V) \mapsto L^2(V_0)$ of a signal on $V$ to $V_0$, which is nothing but the projection of the coordinates to those in $V_0$. We call $G$ the ambient graph. 

\begin{figure}[!htb]
\centering
\includegraphics[width=0.3\textwidth]{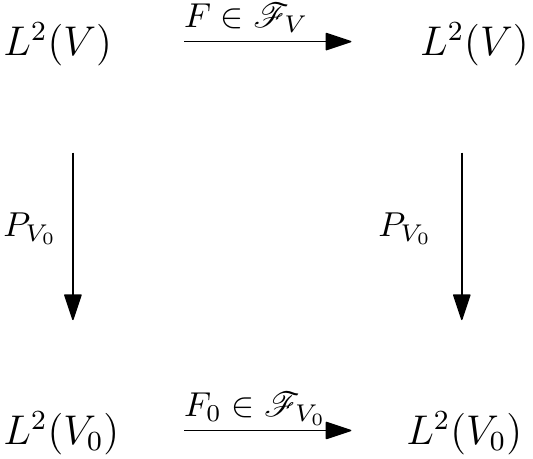}
\caption{The two vertical maps are the projection of a signal on $V$ to that on $V_0$. The two horizontal maps $F$ and $F_0$ are chosen from the prescribed spaces of transformations $\scF_V$ and $\scF_{V_0}$, making the diagram commute for $\ell$ being the operator difference and $\epsilon=0$.}
\label{fig:ssp6}
\end{figure}

\begin{Definition} \label{defn:sma}
Suppose $\scF_V$ and $\scF_{V_0}$ are fixed sets of filters on $L^2(V)$ and $L^2(V_0)$ respectively (cf.\ \cref{fig:ssp6}). Furthermore, assume that for each $F_0 \in \scF_{V_0}$, there is an eigenbasis of $L^2(V_0)$ consisting of eigenvectors of $F_0$. For a given loss function $\ell(\cdot,\cdot)$ and constant $\epsilon\geq0$, a \emph{subgraph signal processing (subGSP) transformation pair} \gls{wrt} $(\scF_V,\scF_{V_0})$ is a pair $(F,F_0)$ with $F \in \scF_V$ and $F_0\in \scF_{V_0}$ such that $\ell(P_{V_0}\circ F, F_0\circ P_{V_0})\leq \epsilon$.
\end{Definition}

For example, if $\ell(L_1,L_2)=\norm{L_1-L_2}$ where $\norm{\cdot}$ is the operator norm and $\epsilon=0$, \cref{defn:sma} requires that the pair $(F,F_0)$ satisfies $\PV{0}\circ F = F_0 \circ \PV{0}$. We call this loss $\ell$ the \emph{operator difference}, which is one of the main loss functions used in the sequel. Intuitively, we want $F_0$ to play the role of a shift operator on $L^2(V_0)$. On the other hand, we want it to be related to filters on $L^2(V)$, and this is achieved via the relation $P_{V_0}\circ F= F_0\circ P_{V_0}$.

\begin{Example} \label{eg:lgbt}
Let $G$ be a directed cycle graph (e.g., \cref{fig:subgraph}) and $A_G$ be the adjacency matrix of $G$, whose orthonormal eigenbasis $\{y_k : k=0,\ldots,|V|-1\} $ gives the basis of the DFT. More specifically, $y_k = \ofrac{\sqrt{|V|}}(\exp(\ui 2\pi kj/|V|))_{0\leq j\leq |V|-1}$, where $\ui=\sqrt{-1}$. Let $\lambda_k=\exp(-\ui 2\pi k/|V|)$ be the corresponding eigenvalue.

Let $V_0$ be a subset of equally spaced vertices on $G$. Suppose $|V_0|> 1$ is a proper divisor of $|V|$. Let $m=|V|/|V_0|$. Without loss of generality, suppose the projection $\PV{0}$ extracts only vertices at indices $jm$, for $j=0,\ldots,|V_0|-1$. The set 
\begin{align*}
\braces*{\sqrt{m}\PV{0}(y_k) = \ofrac{\sqrt{|V_0|}}(\exp(\ui 2\pi kj/|V_0|))_{0\leq j\leq |V_0|-1} : k=0,\ldots,|V_0|-1}
\end{align*} 
forms a basis of the DFT on the cycle graph of $|V_0|$ vertices with adjacency matrix $F_0$. Let $F = A_G^{m}$. For each $1\leq k\leq |V|$, we have 
\begin{align*}
\PV{0}\circ F(y_k) 
&= \lambda_i^m\PV{0}(y_k) \\
&= \exp(-\ui 2\pi k/|V_0|) \PV{0}(y_{k\Mod |V_0|}) \\
&= F_0\circ\PV{0}(y_k).
\end{align*} 
Note that in this example, $F_0$ is a shift operator of the smaller cycle graph with $|V_0|$ vertices in the traditional GSP sense \cite{Shu13}. In fact, the identity of the operators $P_{V_0}\circ F = F_0\circ P_{V_0}$ holds without referring to any specific test vectors, which is special in this homogeneous setting. 
\end{Example}

\cref{defn:sma} is vacuous without specifying $\scF_V$ and $\scF_{V_0}$. Specifying a reasonable and sufficiently rich family of filters constitutes a large bulk of our work and is described in detail in the next two sections. 
By our assumption on $\scF_{V_0}$, there is an orthonormal basis $\{u_0,\ldots,u_{|V_0|-1}\}$ of eigenvectors of $F_0$. We may further assume that they are ordered in non-decreasing order according to the magnitude of their associated eigenvalues. We may use $\{u_0,\ldots,u_{|V_0|-1}\}$ as a graph Fourier basis to represent signals on $V_0$. 

The conditions we impose on $\scF_{V_0}$ is mainly algebraic to ensure we can perform signal processing tasks as listed above. The resulting space can have dimension up to $|V_0|$. On the other hand, $\scF_V$ should be geometrically defined by taking into account the topology of the embedding of $V_0$ in $G$. In general, we want $\scF_V$ to be of smaller dimension than $|V_0|$. More specifically, we want the transformation in $\scF_V$ to model shifts on $G$. However, it will be an over-simplification to consider just the usual shift operators such as adjacency or Laplacian matrices as illustrated in the following example. Consider the operator difference loss with $\epsilon=0$. For each signal $x \in L^2(V_0)$, a signal $y\in L^2(V)$ is an extension of $x$ if $P_{V_0}(y)=x$. For each $x$, all its extensions can be identified with $\mathbb{R}^{|V|-|V_0|}$. Requiring $P_{V_0}\circ F = F_0\circ P_{V_0}$ amounts to $P_{V_0}\circ F (y_1) = P_{V_0}\circ F(y_2)$ for any two extensions $y_1$ and $y_2$ of the same $x$. Therefore, if we write $F$ in the matrix form, the bottom left $(|V|-|V_0|)\times |V_0|$ block must be all zeros. This example shows that it is too restrictive to require $F$ being a shift operator in GSP theory, (or even symmetric), or the loss to be zero. 

Now we are in the position to give a preview of our approach to tackle the above issues. Firstly, in \cref{sec:semi}, we introduce new families of filters, called \emph{semi shift invariant} filters. These families are intermediates between the space of shift invariant filters and the space of all linear filters. Moreover, a semi shift invariant filter is defined by piecing together shifts on different parts of the graph. Such a filter serves as a suitable candidate to be included in $\scF_V$, because $V_0$ may be in-homogeneously embedded in $V$. 
Secondly, in actual learning, it is usually too restrictive to require $\ell(P_{V_0}\circ F, F_0\circ P_{V_0})=0$ once we fix parametrized families $\scF_V$ and $\scF_{V_0}$. For example, if $\scF_V$ is parametrized by a small parameter family, it is unlikely that we can find a non-trivial pair $(F,F_0)$ with symmetric $F_0$ so that zero operator difference is achieved. Therefore instead, we seek $(F,F_0)$ that minimizes $\ell(P_{V_0}\circ F, F_0\circ P_{V_0})$ directly or as a Lagrangian penalty in an objective function. 



\section{Semi shift invariant filters} \label{sec:semi}

In this section, we propose a family of filters as our choice of $\scF_V$. Let $L_G$ be a fixed shift operator for $G$, whose matrix form we assume to be normal, i.e., it is diagonalizable by a unitary matrix. The graph Laplacian is the primary example. 
\begin{Definition} \label{defn:lvs}
Let $V_0 \subset V$ and $d\leq |V|-1$. Suppose $Q_d$ is a polynomial of degree $d$. Let $\oPV{0}: L^2(V)\mapsto L^2(V)$, $f \mapsto \oPV{0} f$ be such that $\oPV{0} f(v) = f(v)$ for $v\in V_0$ and $0$ on $V\backslash V_0$. A \emph{semi shift invariant filter} supported on $V_0$ of degree $d$ is a composition $\oPV{0}\circ Q_d(L_G)$. 
\end{Definition}

Intuitively, a semi shift invariant filter of degree $d$ gathers $d$-hop information locally on $V_0$. Therefore, it should be viewed as a local version of a shift invariant filter. 

\begin{Lemma}\label{lem:dhops}
For $V_0 \subset V$ and $d \geq 0$, the $d$-hop neighborhood $B_d(V_0)$ of $V_0$ is the union of vertices that are at most $d$ hops away from some vertex of $V_0$. Suppose $L_G$ is the Laplacian of $G$ and $L_{V_0,d}$ is the Laplacian of the induced subgraph on $B_d(V_0)$, extended by $0$ to $V\backslash B_d(V_0)$. Then a semi shift invariant filter $F = \oPV{0} \circ Q_d(L_G)$ supported on $V_0$ of degree $d$ is also given by $F = \oPV{0}\circ Q_d(L_{V_0,d})$.
\end{Lemma}

\begin{IEEEproof}
Let $x$ be any graph signal. As $Q_d$ is of degree $d$, the value of $Q_d(L_G)(x)$ at each vertex $v\in V$ depends only on the signal values at vertices at most $d$ hops away from $v$. The filters $Q_d(L_G)$ and $Q_d(L_{V_0,d})$ are thus the same on $V_0$. As the signal value of $F(x)$ is $0$ outside $V_0$, we have the desired identity $F = \oPV{0}\circ Q_d(L_{V_0,d})$.  
\end{IEEEproof}

\Cref{lem:dhops} allows a fast computation of a semi shift invariant filter if the size of $B_d(V_0)$ is small. We now extend \cref{defn:lvs} to a collection of subsets.

\begin{Definition}\label{defn:glvs}
For $k\geq 1$, let $C = (V_1,\ldots, V_k)$ be a tuple of subsets of vertices in $V$ and $D = (d_1,\ldots,d_k)$ be a tuple of non-negative integers each smaller than $|V|$. The space of semi shift invariant filters $\scF_{C,D}$ on $C$ of degree type $D$ is the span of semi shift invariant filters supported on $V_j$ of degree $d_j$ for $1\leq j\leq k$, i.e., if $F\in\scF_{C,D}$, then $F = \sum_{j=1}^k \oPV{j} \circ Q_{d_j}(L_G)$ for some polynomials $Q_{d_1},\ldots,Q_{d_k}$.
\end{Definition}

For any $C$ and $D$, $\scF_{C,D}$ is a vector space. Therefore we have the notion of linear independence of elements in $\scF_{C,D}$, which is utilized in some of our discussions below. For special choices of $C$ and $D$, we recover familiar families of filters. For convenience, if $C=\{v\}$ and $D = d$ for some vertex $v$ and non-negative integer $d$, we write $\scF_{C,D}$ as $\scF_{v,d}$.

\begin{Example} \label{eg:ica}
\begin{enumerate}[a)]
\item Suppose $L_G$ does not have repeated eigenvalues. If $C = V$ and $D = |V|-1$, then $\scF_{C,D}$ is the space of all shift invariant filters in the usual GSP theory. By simple linear algebra, each element of $\scF_{C,D}$ is a polynomial of $L_G$ of degree not more than $|V|-1$.

\item \label{it:sld} Suppose $L_G$ does not have repeated eigenvalues, and no eigenvector of $L_G$ has zero components. For each $v\in V$ and $0\leq d\leq |V|-1$, we have $\dim \scF_{v,d}=d+1$. To see this, we note that the space $\scF_{v,d}$ is spanned by $\overline{P}_{v}\circ L_G^i$, for $0\leq i\leq d$. Therefore, $\dim\scF_{v,d} \leq d+1$. We also have $\dim \scF_{v,d+1} - \dim \scF_{v,d} \leq 1$. Hence, it suffices to prove the case for $d=|V|-1$. Suppose 
\begin{align} \label{eq:sl}
\sum_{0\leq i\leq |V|-1} a_i \overline{P}_{v}\circ L_G^i = 0.
\end{align} 
We write the orthonormal decomposition of  $L_G = U\Lambda U^{*}$. The diagonal entries of $\Lambda$ are the distinct eigenvalues $\lambda_0,\ldots, \lambda_{|V|-1}$. Suppose the index of $v$ is $j$. Unwrapping \cref{eq:sl} as a transformation in the frequency domain, we have 
\begin{align*}
\sum_{0\leq i \leq |V|-1}a_i (U_{j0}\lambda_0^i,\ldots, U_{j,|V|-1}\lambda_{|V|-1}^{i}) = 0.
\end{align*}
As the row vectors $(\lambda_0^i,\ldots, \lambda_{|V|-1}^{i})$, $0\leq i\leq |V|-1$ are linearly independent and the numbers $U_{j0}, \ldots, U_{j,|V|-1}$ are non-zero, the vectors $(U_{j0}\lambda_0^i,\ldots, U_{j,|V|-1}\lambda_{|V|-1}^{i})$, $0\leq i\leq |V|-1$ are linearly independent. Consequently, $a_i=0$ for $0\leq i\leq |V|-1$.

\item \label{it:sl} Suppose $L_G$ does not have repeated eigenvalues, and no eigenvector of $L_G$ has zero components. If $C = (\{v\})_{v\in V}$ and $D = (|V|-1,\ldots, |V|-1)$, then $\scF_{C,D}$ is the space of all filters on $L^2(V)$. To see this, we note that the dimension of the space of all linear filters is $|V|^2$. For any vertices $u, v\in V$ and polynomials $Q$ and $Q'$, $\overline{P}_{u}\circ Q(L_G)$ and $\overline{P}_{v}\circ Q'(L_G)$ are supported on different vertices and are thus linearly independent in $\scF_{C,D}$. It suffices to show that for each $v\in V$, $\dim \scF_{v,|V|-1} = |V|$, but this follows from \ref{it:sld} in this example. More generally, if $D = (l,\ldots,l)$ for some $l\leq |V|-1$, then $\scF_{C,D}$ is the space of \emph{node-variant graph filters} up to degree $l$ described in \cite{seg17}. 
\end{enumerate}
\end{Example}

For our applications in the paper, we are also interested in other intermediate cases.

\begin{figure}[!htb]
\centering
\includegraphics[width=0.6\columnwidth]{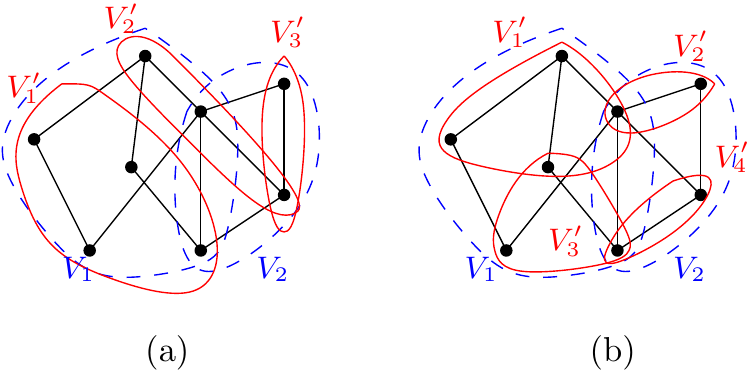}
\caption{In (a), $C= (V_1,V_2)$ (blue) and $C' = (V_1', V_2', V_3')$ (red). $C'$ is not a refinement of $C$ because $V_2'$ is contained in neither $V_1$ nor $V_2$. In (b), $C= (V_1,V_2)$ (blue) and $C' = (V_1', V_2', V_3', V_4')$ (red). Though \cref{def:cic}\ref{it:evi} is satisfied, $C'$ is not a refinement of $C$ because $V_1', V_3'$ are both in $V_1$ but $V_1'\cap V_3' \neq \emptyset$.}
\label{fig:ssp5}
\end{figure}

\begin{Definition} \label{def:cic}
The tuple $C = (V_1,\ldots, V_k)$ is called \emph{essential} if for each $1\leq i\leq k$, \begin{align*} V_i \backslash \bigcup_{1\leq j\neq i\leq k}V_j \neq \emptyset.\end{align*} A tuple $C' = (V_1',\ldots, V_{l}')$ is said to be a \emph{refinement} of $C$ if the following holds (cf.\ \cref{fig:ssp5}):
\begin{enumerate}[i)]
\item $\bigcup_{1\leq i\leq k}V_i = \bigcup_{1\leq j\leq l}V_j'$.
\item \label{it:evi} Each $V_j'$ is contained in some $V_i$.
\item If distinct $V_{j_1}', V_{j_2}'$ are contained in the same $V_i$ for some $i$, then $V_{j_1}'\cap V_{j_2}' = \emptyset$.
\end{enumerate}
\end{Definition}

We now summarize our main structural result on $\scF_{C,D}$ for various tuples $C$ and $D$. We compute the dimension of $\scF_{C,D}$ based on conditions on $C$ and $D$. Moreover, for different $(C,D)$ and $(C',D')$, we discuss inclusion relations between $\scF_{C,D}$ and $\scF_{C',D'}$. This is useful in designing filter banks we may sometimes want to avoid a family of filters to be contained in another. 

\begin{Proposition} \label{thm:ivv}
If $V_1 \subset V_2 \subset V$ and $0\leq d_1 \leq d_2 \leq |V|-1$, then $\dim \scF_{V_1,d_1} \leq \dim \scF_{V_2,d_2}$. Furthermore, suppose $L_G$ does not have repeated eigenvalues, and no eigenvector of $L_G$ has zero components. Then, the following holds.
\begin{enumerate}[a)]
\item\label{it:ivva}  If $C = (V_1,\ldots, V_k)$ is essential and $D = (d_1,\ldots,d_k)$, then $\dim \scF_{C,D} = \sum_{1\leq i\leq k} d_i +k$.

\item Let $C = (V_1,\ldots, V_k)$, $D=(d_1,\ldots,d_k)$, $C' = (V_1',\ldots, V_l')$ and $D' = (d_1',\ldots, d_l')$. Suppose $C'$ is a refinement of $C$. If $D$ and $D'$ satisfy $d_i\leq d_j'$ whenever $V_j'\subset V_i$, then $\scF_{C,D} \subset \scF_{C',D'}$. On the other hand, suppose that $\scF_{C,D} \subset \scF_{C',D'}$ and $\bigcup_{1\leq j \leq l} V_j'\subset \bigcup_{1\leq j \leq k} V_j$. If $D$ and $D'$ are constant tuples with $d_i=d_l'=d$ for all $i, l$ and the same $0\leq d <n$, and $C'$ is essential, then $C'$ is a refinement of $C$. 
\end{enumerate}
\end{Proposition}

\begin{IEEEproof}
As $V_1\subset V_2$ and $d_1\leq d_2$, any $F\in \scF_{V_1,d_1}$ is the projection $\oPV{1}\circ F'$ of some $F' \in \scF_{V_2,d_2}$ to $V_1$. Furthermore, if $F_1',\ldots, F_k' \in \scF_{V_2,d_2}$ are filters such that $\oPV{1}\circ F_1',\ldots, \oPV{1}\circ F_k'$ are linearly independent in $\scF_{V_1,d_1}$, then $F_1',\ldots, F_k'$ are linearly independent in $\scF_{V_2,d_2}$ since the sum and projection operations commute. Hence, a basis of $\scF_{V_1,d_1}$ consists of the projection of linearly independent filters in $\scF_{V_2,d_2}$, whence $\dim \scF_{V_1,d_1}\leq \dim \scF_{V_2,d_2}$. We next further assume that $L_G$ does not have repeated eigenvalues, and no eigenvector of $L_G$ has zero components.

\begin{enumerate}[a)]
\item As $C$ is essential, each $V_i$ contains a vertex $v_i$ outside every $V_j, j\neq i$. By \cref{eg:ica}\cref{it:sld} and the above result, we have $d_i+1\leq \dim \scF_{v_i,d_i} \leq \dim \scF_{V_i,d_i} \leq d_i+1$. Therefore, we must have $\dim \scF_{v_i,d_i} = \dim \scF_{V_i,d_i}$, and the (surjective) projection $\overline{P}_{v_i}: \scF_{V_i,d_i} \to \scF_{v_i,d_i}$ is an isomorphism.

For each $i$, let $C_{-i}$ be obtained from $C$ by removing $V_i$ and $D_{-i}$ be obtained from $D$ by removing $d_i$. If $F \in \scF_{v_i,d_i}$ and $F'\in \scF_{C_{-i},D_{-i}}$ where both $F$ and $F'$ are not trivially $0$, then they are linearly independent of each other. Indeed, if $aF+bF'=0$, then applying $\overline{P}_{v_i}$, we have $a\overline{P}_{v_i}\circ F=0 \in \scF_{v_i,d_i}$. As $\overline{P}_{v_i}\circ F$ is non-zero, we must have $a=0$ and hence $b=0$. Consequently, filters in $\scF_{V_i,d_i}$, for $1\leq i \leq k$, are all linearly independent and $\dim \scF_{C,D} = \sum_{1\leq i\leq k}\scF_{V_i,d_i} = \sum_{1\leq i\leq k}d_i+k$.

\item If $C'$ is a refinement of $C$, then each $V_i$ in $C$ is a disjoint union of $V_i = \bigcup_{j=1}^{k_i} V_{i_j}'$ with each $V_{i_j}'$ in $C'$. Let $F = \overline{P}_{V_i}\circ \sum_{0\leq j \leq d_i}a_j L_G^j \in \scF_{V_i,d_i}$, where $a_j$, $0\leq j\leq d_i$, are scalars. Then since $d_i \leq d_{i_j}'$, we have 
\begin{align*}
F = \sum_{1\leq j\leq k_i} \parens*{ \overline{P}_{V_{i_j}'}\circ \sum_{0\leq j \leq d_i}a_jL_G^j} \in \scF_{C',D'}.
\end{align*} 

To prove the second claim, we verify each of the conditions in \cref{def:cic}. To show the first condition, suppose $\bigcup_{1\leq i\leq k}V_i$ is not contained in $\bigcup_{1\leq j\leq l}V_j'$. Let $v \in \bigcup_{1\leq i\leq k}V_i \backslash \bigcup_{1\leq j\leq l}V_j'$, and $F \in \scF_{C,D}$ be a filter such that $P_{v}\circ F$ is non-trivial. Such an $F$ exists as $\dim \scF_{v,d} \geq 1$ by \cref{eg:ica}\cref{it:sld}. However, $F \notin \scF_{C',D'}$ as the projection of any filter of $\scF_{C',D'}$ to $v$ is trivial. This gives rise to a contradiction.

For the second condition in \cref{def:cic}, we first note that since the eigenvectors of $L_G$ have no zero components, if $F=\sum_{0\leq j\leq d}a_jL_G^j \ne 0$, then $\overline{P}_v \circ F \ne 0$ for all $v\in V$. Suppose without loss of generality that $V_1'$ is not contained in any single $V_i$. As we assume that $C'$ is essential, there is a $v$ contained only in $V_1'$ and not any other $V_j'$, $j\ne1$. Let $V_i$ in $C$ contain $v$, $V_1'\backslash V_i \ne \emptyset$, and $F = \overline{P}_{V_i}\circ \sum_{0\leq j\leq d}a_jL_G^j \in \scF_{C,D}$ be a non-zero filter. Since $\scF_{C,D}\subset\scF_{C',D'}$, $F \in \scF_{C',D'}$, and by considering the projection to $v$, $F$ must have a summand $F_1 = \overline{P}_{V_1'}\circ \sum_{0\leq j\leq d}a_jL_G^j$. However, the projection of $F_1$ to $V_1'\backslash V_i$ is non-zero. For each $v' \in V_1'\backslash V_i$, there must be some other $V_j'$ such that 
\begin{enumerate}[i)]
\item $v'\in V_1' \cap V_j'$, and
\item $F$ has a non-zero summand $F_j \in \scF_{V_j',d}$.
\end{enumerate}
For such a $V_j'$, there is a $v_j$ contained exclusively (\gls{wrt} $C'$) in $V_j'$. However, $\overline{P}_{v_j}\circ F_j \neq 0$ and hence $v_j \in V_i$. This implies that $F_j = \overline{P}_{V_j'} \circ \sum_{0\leq j\leq d}a_jL_G^j$ since $F_j$ and $F$ have the same projection to $v_j$. In conclusion, for any $v' \in V_1'\backslash V_i$, there is a positive integer $m$ such that $0 = \overline{P}_{{v'}}\circ F = m\overline{P}_{v'}\circ \sum_{0\leq j\leq d}a_jL_G^j \neq 0$, which is a contradiction.

For the last condition in \cref{def:cic}, consider any $V_i$ and choose any non-zero filter 
\begin{align*}
F \in \scF_{V_i,d} \subset \scF_{C,D} \subset \scF_{C',D'}.
\end{align*}
For any $V_j'\subset V_i$, there is a $v_j$ contained exclusively (\gls{wrt} $C'$) in $V_j'$. Therefore, $F$ has a summand $\overline{P}_{V_j'}\circ F$. Then for any $v\in V_i$, $\overline{P}_{v}\circ F$ is the same as $m\overline{P}_{v}\circ F$, where $m$ is the number of $V_j'\subset V_i$ that contains $v$. Hence, $m=1$. Therefore, for any distinct $V_{j_1}', V_{j_2}' \subset V_i$, we have $V_{j_1}'\cap V_{j_2}' =  \emptyset$. The proof that $C'$ is a refinement of $C$ is now complete.
\end{enumerate}
\end{IEEEproof}

The family $\scF_{C,D}$ performs different degrees of shifts on different parts of the graph. This is exactly what we want when $V_0$ is non-homogeneously embedded in $G$. For suitably chosen $C$ and $D$ based on the geometry of $V$ and $V_0$ as detailed in \cref{sec:xv}, $\scF_{C,D}$ is our choice of $\scF_V$ in \cref{defn:sma}. 

To end this section, we briefly describe how to express a semi shift invariant filter $F=\oPV{0}\circ Q_d(L_G) \in \FV{0,d}$ supported on $V_0\subset V$, in the frequency domain of the shift operator $L_G$. Let $y_0,\ldots, y_{|V|-1}$ be an eigenbasis consisting of orthonormal eigenvectors of $L_G$ with corresponding eigenvalues $\lambda_0,\ldots, \lambda_{|V|-1}$. Then for a signal $f = \sum_{0\leq i\leq |V|-1}\hat{f}_i y_i$, where $\hat{f}_i=\ip{f}{y_i}$ is its $i$-th GFT coefficient, the vertex $v$ component of $F(f)\in\Real^{|V|}$ is given by 
\begin{align} \label{eq:tjc}
\left\{ 
\begin{array}{rl} 
\sum_{0\leq i\leq |V|-1}Q_d(\lambda_i) \hat{f}_i y_i(v), & \text{if } v\in V_0, \\ 
0, &\text{ otherwise.}
\end{array} \right.
\end{align}
In general, a semi shift invariant filter is a linear combination $F = \sum_{j=1}^k \oPV{j} \circ Q_{d_j}(L_G)$ for some subsets $V_1,\ldots,V_k$, polynomials $Q_{d_1},\ldots,Q_{d_k}$ and positive integer $k$. Therefore, the vertex $v$ component of $\hat{f}_iy_i$ is transformed to $\lambda_{i,v}\hat{f}_iy_i(v)$, where 
\begin{align}\label{eq:lambdaiv}
\lambda_{i,v} = \sum_{\substack{1\leq j \leq k,\\ v\in V_j}} Q_{d_j}(\lambda_i)
\end{align}
is a polynomial of $\lambda_i$. 

For a subGSP transformation pair $(F,F_0)$, to relate their spectrums,  there is a subtlety. In general, $F$ is not symmetric, and hence its eigenvalues might be complex numbers. Instead of considering the eigenvalues of $F$, the above discussions suggest to look at the behavior of $F$ at each $v \in V_0$. More specifically, for each $v\in V_0$, consider $\lambda_{i,v}$ given by \cref{eq:lambdaiv}. We define $\Lambda_v = (\lambda_{i,v})_{0\leq i\leq |V|-1}$, as an ordered tuple. There can be different $v_1,v_2 \in V$ such that $\Lambda_{v_1}=\Lambda_{v_2}$ as ordered sets. 
\begin{Definition} \label{defn:tls}
The largest subset of $v \in V_0$ with the same $\Lambda_v$ is called the \emph{main spectral set of $F$} and the set of vertices the \emph{main component of $F$}.
\end{Definition}
In \cref{sec:spectral}, we study how the spectrum of $F_0$ relates to the main spectral set of $F$.

\section{Elements of subgraph signal processing} \label{sec:sga}
In this section, we first describe other approaches for subgraph signal processing and discuss their insufficiency. We then propose an approach based on the tools introduced in \cref{sec:semi}. 

Recall that $G=(V,E)$ is the underlying graph, which we now assume to be connected for convenience. For the most direct and natural approach, let $H_0=(V_0,E_0)$ be the induced subgraph of $V_0$, i.e., $V_0$ is the vertex set of $H_0$ and $u,v \in V_0$ are connected by an edge in $E_0$ if and only if they are connected by an edge in $E$. Let $L_{H_0}$ be $H_0$'s graph Laplacian. It is natural to consider $L_{H_0}$ and its polynomials for $\scF_{V}$ as in \cref{defn:sma}. 

\begin{figure}[!htb]
\centering
\includegraphics[scale=1.4]{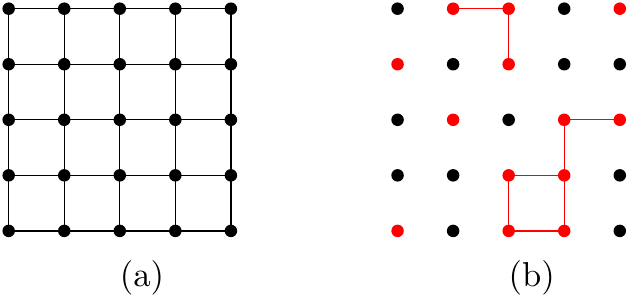}
\caption{Suppose $G$ is the $5\times 5$ lattice in (a). If we choose $p=1/2$, an example of $H_0$ is shown in (b) with $|V_0|=13$, highlighted in red. We see that the largest component of $H_0$ has $6$ vertices and $6$ connected components.}
\label{fig:ssp1}
\end{figure}

To motivate the need for an improvement over this choice of $\scF_V$, consider the simple random model of including a vertex from $V$ in $V_0$ independently with probability $p$. The expected size of $V_0$ is thus $p|V|$. An edge $e$ of $G$ remains in $H_0$ if and only if both ends are in $V_0$, and thus the probability of such an event is $q =p^2$. We want to argue heuristically that with high probability $H_0$ cannot have a large component and hence is highly disconnected if $p$ is small. For this, we compare the random vertex selection model with the Erd\H{o}s-Renyi random model $\mathrm{ER}(q)$ that keeps an edge of $G$ with probability $q$, which is independent over all the edges. The resulting graph $\mathrm{ER}(q)$ thus has the same expected number of edges as $H_0$. However, the edges in $H_0$ tend to cluster together as only edges not sharing a vertex are retained independent of each other. Therefore, if with high probability $\mathrm{ER}(q)$ only has small components, then so does $H_0$. On the other hand, it is known (see \cite{Fri04, Fan09}) for a lot of cases that if $q$ is small enough, then $\mathrm{ER}(q)$ and hence $H_0$ tends to have small components, and becomes highly disconnected (an example is shown in \cref{fig:ssp1}). We formalize parts of this heuristic argument in \cref{app:con}.

Suppose a graph signal $y$ on $G$ is only observed at $V_0$ as $x = P_{V_0}(y)$. If the induced subgraph $H_0$ does not capture enough topological properties of $G$ (e.g., when $H_0$ is highly disconnected with small components), then it can be erroneous to perform signal processing on $x$ with a shift operator of $H_0$ as the signals tend to concentrate in a small part of the graph. The framework introduced in \cref{sec:sub} is designed to overcome this difficulty. 

On the other hand, another useful approach for graph reduction is the Kron reduction \cite{Dor13}: for a given proper subset of vertices $V_0$, the Kron reduced graph operator represented as a matrix is given by
\begin{align}
K = L_{G, V_0V_0} - L_{G, V_0 V\backslash V_0} L_{G, V\backslash V_0 V\backslash V_0}^{-1} L_{G, V\backslash V_0 V_0},
\end{align}
where $L_{G, IJ}$ denotes the submatrix of $L_G$ with rows from the index set $I$ and columns from the index set $J$. For example, it reduces a cycle graph with $8$ nodes to a clique with $4$ nodes. Kron reduction enjoys a few topological properties \cite[Theorem 3.4]{Dor13} and spectral properties \cite[Theorem 3.5]{Dor13}. For example, two nodes are connected in the Kron reduced graph if and only if they are connected by a path in $G$. Its spectrum satisfis the interlacing property, which gives a qualitative bound of the eigenvalues of the Kron reduced operator by the eigenvalues of the original graph shift operator. However, for signal processing purposes, it is more desirable if there are quantitative description of the spectrum of the operators on the ambient graph and subgraph. 

There are some other shortcomings of Kron reduction. It is less flexible as compared to our framework as we do not have specific requirement on the base shift operator $L_G$ on $G$. We have seen it can be the usual Laplacian or the adjacency matrix in the directed case. Moreover, it can also be a polynomial or the normalized version of the above candidates. However, the Kron reduction usually does not respect polynomial filter construction. Geometrically, we will see in \cref{eg:lvb} that if we apply the subGSP framework on the cycle graph with an evenly spaced subset of vertices $V_0$, we obtain a smaller cycle graph on $V_0$, which agrees with the DFT theory (cf. \figref{fig:subgraph}). However, the Kron reduction produces a clique and overdoes with edge connections.

To complete the picture envisioned in \cref{defn:sma}, we need to describe $\scF_{V_0}$ and $\scF_V$.

\subsection{Choice of \texorpdfstring{$\scF_{V_0}$}{FV0}} \label{sec:xv0}

In this subsection, we outline a few candidates for $\scF_{V_0}$. We start with a preliminary definition. 

\begin{Definition}\label{def:H0}
Let $V_0 \subset V$ be a subset of vertices and $H_0=(V_0,E_0)$ be the induced subgraph of $V_0$ in $G$. An \emph{extension} $H$ of $H_0$ is a weighted graph with vertex set $V_0$ and weight function $w(\cdot,\cdot)$ such that: for any pair of vertices $u,v \in V_0$ connected by an edge in $H_0$ (and hence in $G$), $w(u,v)$ is the same as its weight in $H_0$. We denote the set of all such $H$ by $\calE_{H_0}$. 
\end{Definition}

From the definition, the set $\calE_{H_0}$ of all $H$ extending $H_0$ is parametrized by non-negative real numbers each associated with a pair of vertices $u,v \in V_0$ that are not connected by an edge in $H_0$. For example, if $H_0$ is an undirected graph, as a real manifold, the dimension of $\calE_{H_0}$ is $|V_0|(|V_0|-1)/2-|E_0|$.

We now list a few candidates for $\scF_{V_0}$, ordered by set inclusion.
\begin{enumerate}[a)]
\item  \label{it:xtl}$\scF_{V_0} = \{ \text{the Laplacian of a graph in } \calE_{H_0}\}$. Any $F_0 \in \scF_{V_0}$ captures partial geometric features of the embedding of $V_0$ in $G$. However, the downside is that this space can be less flexible due to the small degree of freedom if $H_0$ is well connected with large $|E_0|$.  

\item \label{it:xtlo}$\scF_{V_0} = \{ \text{the Laplacian of any graph of size } |V_0|\}$. There is no particular restriction on $F_0 \in \scF_{V_0}$ other than requiring it to be geometric, i.e., being a graph Laplacian. The Kron reduced Laplacian is included in this family. In \cref{sec:sim}, we see that the optimal $F_0$ for specific applications often do not correspond to the Kron reduced Laplacian.

\item \label{it:xym} $\scF_{V_0} = \{\text{symmetric $|V_0|\times |V_0|$ matrices whose rows sum to }0\}$. This space is large enough as it includes not only graph Laplacians but also their shift invariant families. We require the rows of such $F_0$ sum to $0$, so that constant vectors belong to the $0$-eigenspace, i.e., constant vectors are the smoothest. 
\end{enumerate}

These are our choices of $\scF_{V_0}$ for most applications. One may make other choices if the situation changes such as when the graph is directed. We note that each of the choice gives rise to a submanifold of the space of $|V_0|\times |V_0|$ matrices. Such a manifold is parametrized by $|V_0|^2$ matrix entries subject to respective constraints on the entries.

\subsection{Choice of \texorpdfstring{$\scF_{V}$}{FV}} \label{sec:xv}

For $\scF_V$, we want to make full use of the geometry of the embedding of $V_0$ in $G$. For this, we propose $\scF_V=\scF_{C_{V_0},D_{V_0}}$ to be a semi shift invariant family (cf. \cref{defn:lvs}) with appropriately chosen $C_{V_0}$ and $D_{V_0}$. 

\begin{figure}[!htb]
\centering
\includegraphics[width=0.4\linewidth]{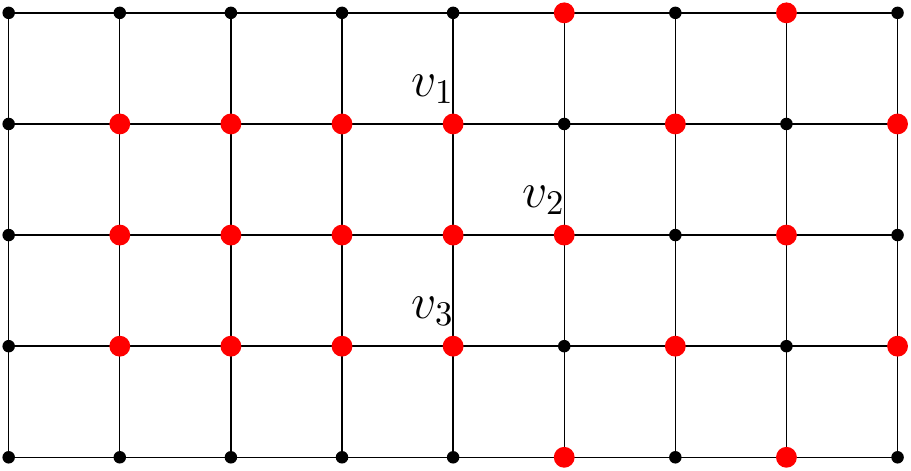}
\caption{Suppose $G$ is the $5\times 9$ lattice graph and $V_0$ consists of the 22 red vertices. The tuple $C_{V_0}= (V_1, V_2)$, where $V_1$ consists of the $13$ vertices, including $v_1,v_2,v_3$ and all the red vertices on their left. $V_2$ has $12$ vertices, including $v_1,v_2,v_3$ and all the red vertices on their right. If $r = 1$, then $D_{V_0} = (2,3)$. All other $V_i$, $i\geq 3$ are empty.}
\label{fig:ssp7}
\end{figure}

Consider two distinct vertices $v_1, v_2 \in V_0$ being $d$ hops away from each other in $G$. In order to exchange signals between $v_1$ and $v_2$ in $G$, the $d$-th power of the Laplacian $L_G$ is needed. Therefore, locally at each $v \in V_0$, the filter $F$ in \cref{fig:ssp6} should take the form of a polynomial in $L_G$, whose degree depends on how far it is away from other vertices in $V_0$. In view of this, we define $C_{V_0}$ and $D_{V_0}$ as follows. An illustration is given in \cref{fig:ssp7}.

\begin{Definition} \label{defn:gas}
Let $\delta_G$ be the diameter of $G$. Given a subset of vertices $V_0 \in V$, let $\{V_i \mid 1\leq i\leq \delta_G\}$ be a collection of subsets of $V_0$ such that the following holds:
\begin{enumerate}[i)]
\item \label{it:bv} $\bigcup_{1\leq i\leq \delta_G}V_i = V_0$. 
\item \label{it:fei} The set $V_1$ consists of those vertices in $V_0$ that have 1-hop neighbors in $V_0$ as well as these neighbors. For each $i=2,\ldots,\delta_G$, if $v\in V_0$ has $i$-hop (calculated in $G$) neighbors in $V_0$ but no $j$-hop neighbors for $j\leq i-1$ in $V_0$, then $v$ and all of its $i$-hop neighbors are in $V_i$.  
\end{enumerate}
Fix a small integer $r\geq 0$. Let $\calI_{V_0}$ be the set of $i=1,\ldots,\delta_G$ such that $V_i$ is non-empty. We include each non-empty $V_i$ in the tuple $C_{V_0}$ and associate it with degree $i+r$ in forming $D_{V_0}$.
\end{Definition}

We now give some intuitions to \cref{defn:gas}. For \ref{it:bv} in \cref{defn:gas}, we want the family of filters to perform local operations on $V_0$, therefore we require the sets $\{V_i \mid i\in \calI_{V_0}\}$ cover $V_0$. For \ref{it:fei}, each vertex in $V_i$ finds at least an $i$-hop neighbor. Therefore, to transfer signals in $V_i$, we need to include the $i$-th power of the graph shift operator as discussed earlier. To allow flexibility, we may relax the degree to $i+r$ on each $V_i$ with an integer hyper-parameter $r\geq 0$ to allow tuning and adjustment. In summary, a $F\in\scF_{C_{V_0},D_{V_0}}$ has the form
\begin{align}\label{eq:F}
	F = \sum_{i\in\calI_{V_0}} \oPV{i}\circ Q_{i+r}(L_G) = \sum_{i\in\calI_{V_0}}\sum_{j=0}^{i+r} a_{ij} \oPV{i} L_G^j.
\end{align}

The collection $C_{V_0}$ is not necessarily essential (c.f.\ \cref{def:cic}). To obtain an essential tuple, we may take refinements of $C_{V_0}$ in \cref{defn:gas}. By \cref{thm:ivv}\ref{it:ivva}, the dimension of $\scF_{C_{V_0},D_{V_0}}$ is then $\calO(\delta_G^2)$,\footnote{$\calO(\cdot)$ stands for the big-O notation.} independent of the choice of $V_0$. In actual applications, we may include additional constraints such as specifying certain polynomial coefficients (c.f.\ \cref{defn:lvs}). 

\subsection{Transformation Pairs for subGSP}

We wish to find an optimal subGSP transformation pair $F\in \scF_V, F_0\in \scF_{V_0}$ such that $\ell(P_{V_0}\circ F,F_0\circ P_{V_0})$ is minimized. We consider the following problem. Throughout, we assume that $\FV{0}$ and $\FV{}$ are one of the choices proposed in \cref{sec:xv0} and \cref{sec:xv}, respectively, and that the optimal solution exists.

\begin{prob} \label{prob:thh}
\begin{align}
(F^*,F^*_0)= \argmin_{F\in \scF_V, F_0\in \scF_{V_0}} \ell\parens*{P_{V_0}\circ F, F_0\circ P_{V_0}}.
\end{align}
\end{prob}
Additional regularization or conditions might be imposed to prevent trivial solutions. As an example, with the choice $\scF_V =\scF_{C_{V_0},D_{V_0}}$ described in \cref{sec:xv}, we can fix the coefficient $a_{11}$ of the $\oPV{1}L_G$ term in \cref{eq:F} to be the constant 1.

\begin{figure}[!htb]
\centering
\includegraphics[width=.35\linewidth]{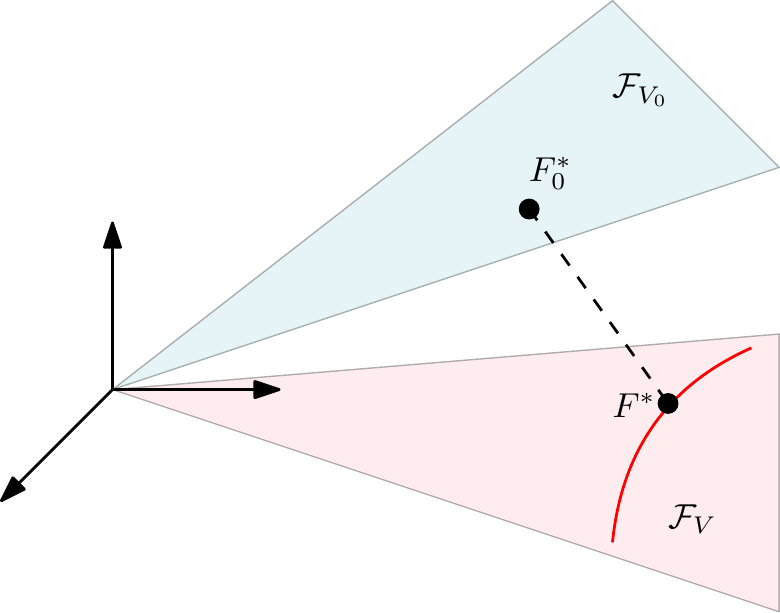}
\caption{The optimization in \cref{prob:thh} using the operator difference loss seeks to find two filters belonging to sparse spaces of operators that are close to each other in the operator norm.}
\label{fig:ssp16}
\end{figure}

Before proceeding further, we give some intuitions for the case where $\ell$ is the operator difference loss. In the space of filters from $\mathbb{R}^{|V|} \to \mathbb{R}^{|V_0|}$ and assuming that $|V_0|$ is much smaller than $|V|$, we have two sparse subspaces $\mathcal{F}_{V} := \{P_{V_0}\circ F \mid F\in \scF_{V}\}$ and $\mathcal{F}_{V_0}: = \{F_0\circ P_{V_0} \mid F_0\in \scF_{V_0}\}$ if we adopt the choices of $\scF_V$ and $\scF_{V_0}$ proposed in the previous subsections. The former contains a family of operators that capture signal correlations on $V_0$, modeled on the signal interactions on $V$ in view of the embedding of $V_0$ in $V$. The latter space $\mathcal{F}_{V_0}$ poses algebraic restrictions on the filters. Due to sparsity, these two spaces hardly intersect non-trivially. Therefore, the optimization \cref{prob:thh} is seeking for $F^*$ and $F^*_0$ from $\scF_v$ and $\scF_{V_0}$ respectively, so that $P_{V_0}\circ F^{*}$ and $F^*_0\circ P_{V_0}$ are closest to each other (as illustrated in \cref{fig:ssp16}). We may understand that for the pair $(F^*, F^*_0)$, $F^*$ models how signals are transferred to $V_0$ when viewed as a subset of $V$, and $F^*_0$ is the actual operator on $V_0$ approximating $F^*$ without referring to $V\backslash V_0$. 

\begin{Lemma}\label{lem:sia}
\begin{enumerate}[a)]
\item Suppose $C'=(V_1',\ldots, V_l')$ is a refinement of $C$. If $D = (d_1,\ldots, d_k)$ and $D' = (d_1',\ldots, d_l')$ satisfy $d_i\leq d_j'$ provided $V_j'\subset V_i$, then for any $\FV{0}$, 
\begin{align*}
\min_{F \in \scF_{C', D'}, F_0 \in \scF_{V_0}} \ell(P_{V_0}\circ F, F_0\circ P_{V_0}) \leq \min_{F \in \scF_{C, D}, F_0 \in \scF_{V_0}} \ell(P_{V_0}\circ F, F_0\circ P_{V_0}).
\end{align*}

\item \label{it:fca} Suppose $F^*= \oPV{0} \circ Q(L_G)$ in solving \cref{prob:thh} with the operator difference loss. For $0\leq i\leq |V|-1$, denote the $i$-th eigenvalue and corresponding orthonormal eigenvector of $L_G$ by $\lambda_i$ and $y_i$, respectively. Let the eigenvalues of $F^*_0$ be $\mu_j$, for $ 0\leq j\leq |V_0|-1$. Then 
\begin{align*}
\max_{0\leq i\leq |V|-1} \min_{0\leq j\leq |V_0|-1} |\mu_j - Q(\lambda_i)|\cdot \norm{P_{V_0}(y_i)}_2 \leq \min_{F \in \scF_{V}, F_0 \in \scF_{V_0}} \norm{P_{V_0}\circ F - F_0\circ P_{V_0} }.
\end{align*}
\end{enumerate}	
\end{Lemma}
\begin{IEEEproof}
\begin{enumerate}[a)]
\item Under the given conditions, we have $\scF_{C',D'} \subset \scF_{C,D}$ by \cref{thm:ivv}. Consequently, we have the inequality
\begin{align*}
\min_{F \in \scF_{C', D'}, F_0 \in \scF_{V_0}} \ell(P_{V_0}\circ F, F_0\circ P_{V_0}) \leq \min_{F \in \scF_{C, D}, F_0 \in \scF_{V_0}} \ell(P_{V_0}\circ F, F_0\circ P_{V_0}).
\end{align*}

\item Let $\{x_j \mid 0\leq j\leq |V_0|-1\}$ be an orthonormal eigenbasis of $F^*_0$. For each $y_i$, write $P_{V_0}(y_i) = \sum_{0\leq j\leq |V_0|-1} a_{ij}x_j$ for some coefficients $a_{ij}$. Therefore, $F^*_0\circ P_{V_0}(y_i) = \sum_{0\leq i\leq |V_0|-1} a_{ij}\mu_j x_j$.  On the other hand, 
\begin{align*}
P_{V_0}\circ F^*(y_i) = P_{V_0}(Q(\lambda_i)y_i) = \sum_{0\leq j\leq |V_0|-1} a_{ij}Q(\lambda_i)x_j.
\end{align*}
Hence, 
\begin{align*}
\norm*{(F^*_0\circ P_{V_0}-P_{V_0}\circ F^*)(y_i)}_2^2 
& = \norm*{\sum_{0\leq j\leq |V_0|-1} a_{ij}(\mu_j-Q(\lambda_i))x_j}_2^2 \\ 
& = \sum_{0\leq j\leq |V_0|-1} \abs*{a_{ij}(\mu_j-Q(\lambda_i))}^2 \\
& \geq \min_{0\leq j\leq |V_0|-1}|\mu_j-Q(\lambda_i)|^2\sum_{0\leq j\leq |V_0|-1} a_{ij}^2\\
& = \min_{0\leq j\leq |V_0|-1} |\mu_j-Q(\lambda_i)|^2\norm*{P_{V_0}(y_i)}_2^2.
\end{align*}
From the definition of the operator norm, we have 
\begin{align*}
\max_{0\leq i\leq |V|-1}\norm*{(F^*_0\circ P_{V_0}-P_{V_0}\circ F^*)(y_i)}_2^2 \leq  \norm{P_{V_0}\circ F^* - F^*_0\circ P_{V_0} }^2,
\end{align*}
and the result follows.
\end{enumerate}
\end{IEEEproof}

\cref{lem:sia}\ref{it:fca} describes how close the two sets of values $\{\mu_j \mid 0\leq j\leq |V_0|-1\}$ and $\{Q(\lambda_i) \mid 0\leq i\leq |V|\}$ are, in terms of the norm of the operator $P_{V_0}\circ F - F_0\circ P_{V_0} $. We shall illustrate with examples later in \cref{sec:sim}. 

The parameter space of $\FV{}$ is usually small, as it is a restricted space of $\calO(\delta_G^2)$ polynomial coefficients. However, that of $\scF_{V_0}$ can be large, with the number of parameters usually of the order $\calO(|V_0|^2)$. As shown in the result below, at the cost of a small increase in operator difference loss, it is possible to reduce the size of the parameter space of $\FV{0}$. 

\begin{Proposition} \label{prop:gva}
Given $V_0$ and its induced subgraph $H_0$, assume that $\beta_{V_0}$ is the minimal operator difference loss obtained by solving \cref{prob:thh} with optimal solution $(F^*, F^*_0) \in \scF\times\scF_{V_0}$. Let $r_{\max}$ be the largest $L^1$-norm of the rows of $F^*_0$. Consider $\scF_{V_0}$ as a space parametrized by $|V_0|^2$ variables corresponding to matrix entries (cf. \cref{sec:xv0}). For any $k>0$, let $\scF_{V_0,k}$ be the subspace of $\scF_{V_0}$ with at most $k$ nonzero parameters. Then, for any $\epsilon>0$, there is a number $N$ of order $\calO(|V_0|/\epsilon^2)$ such that the value
\begin{align} \label{prob:ha}
\beta_{V_0,N} = \min_{F\in \scF_V, F_0\in \scF_{V_0,N}} \norm{P_{V_0}\circ F - F_0\circ P_{V_0}},
\end{align}
satisfies 
\begin{align}\label{prop:ineq}
	\beta_{V_0}\leq \beta_{V_0,N}\leq \beta_{V_0}+\epsilon r_{\max}.
\end{align}
\end{Proposition}
\begin{IEEEproof}
See \cref{app:spa} for the proof and further discussions.
\end{IEEEproof}

\subsection{The Operator Difference Loss and the Least Squares Loss}
In this section, we discuss special cases and different variations of \cref{prob:thh}. We first consider $\ell$ being the operator difference loss. We choose $\FV{0}$ to be the family of $|V_0|\times |V_0|$ positive semi-definite matrices and $\scF_V$ to be that in \cref{defn:gas}. Then, \cref{prob:thh} can be recast as a semi-definite program \cite{Lov03}:
\begin{align}\label[prob]{opt:thh}
\begin{aligned}
	\min &\ t \\
	\st 
	&\ \begin{bmatrix} tI & Z \\ Z\T & tI \end{bmatrix} \succeq 0, \\
	&\ Z = F_0\PV{0} - \PV{0}F, \\
	&\ t \geq 0,\ F_0 \succeq 0,\ \cref{eq:F},
\end{aligned}
\end{align}
where $I$ is the identity matrix. We can also include additional constraints on the polynomial coefficients in \cref{eq:F} by fixing some of them to be known constants.

Next we consider a $L^2$-type loss. A special case of \cref{prob:thh} is the following, where we minimize the total squared loss over a set $\Omega$:
\begin{align*}
\min_{F\in \scF_V, F_0 \in \scF_{V_0}} \int_{(x,y)\in\Omega} \norm{F_0(x)-P_{V_0}\circ F(y)}_2^2 \;\ud\mu(x,y).
\end{align*}
Here, each $y\in L^2(V)$ with the corresponding $x=\PV{0}(y)$, and $\mu$ is a measure. We are interested in sets $\Omega$ that capture the variation of the signal pair $(x,y)$. For example, $\Omega$ should contain $x$ and $y$ of similar levels of smoothness: if $x$ is constant, it is more likely that $y$ is also constant. In this problem formulation, we assume that $L_G$ is the graph Laplacian and study the total squared loss over an $\Omega$ constructed using the following two steps:
\begin{enumerate}[(i)]
\item Take all the pairs $\{(x_i,y_i): 0\leq i\leq |V|-1\}$, where $\{y_i : 0\leq i\leq |V|-1\}$ forms an orthonormal eigenbasis of $L_G$ and $x_i = P_{V_0}(y_i)$. The pairs are ordered according to the magnitude of the eigenvalues corresponding to $y_i$. These eigenvectors $y_i$ are a good representation of signals on $G$, as they are of different degrees of smoothness.

\item \label[step]{it:lzb} Let $0\leq \delta\leq 1$ be a fixed number. We start by including the pair of constant vectors $(x_0,y_0)$ in $\Omega$. Inductively, for $(x_i,y_i)$, we include it in $\Omega$ if the cosine distance between $x_i$ and $x$ for each $(x,y) \in \Omega$ is at most $\delta$. The consideration here is that for pairs $(x_i,y_i)$ and $(x_j,y_j)$, $i\leq j$, if $x_i$ and $x_j$ are close enough in cosine distance, then we prefer excluding $(x_j, y_j)$ from $\Omega$. 
\end{enumerate}

\begin{figure}[!htb]
\centering
\includegraphics[scale=0.7]{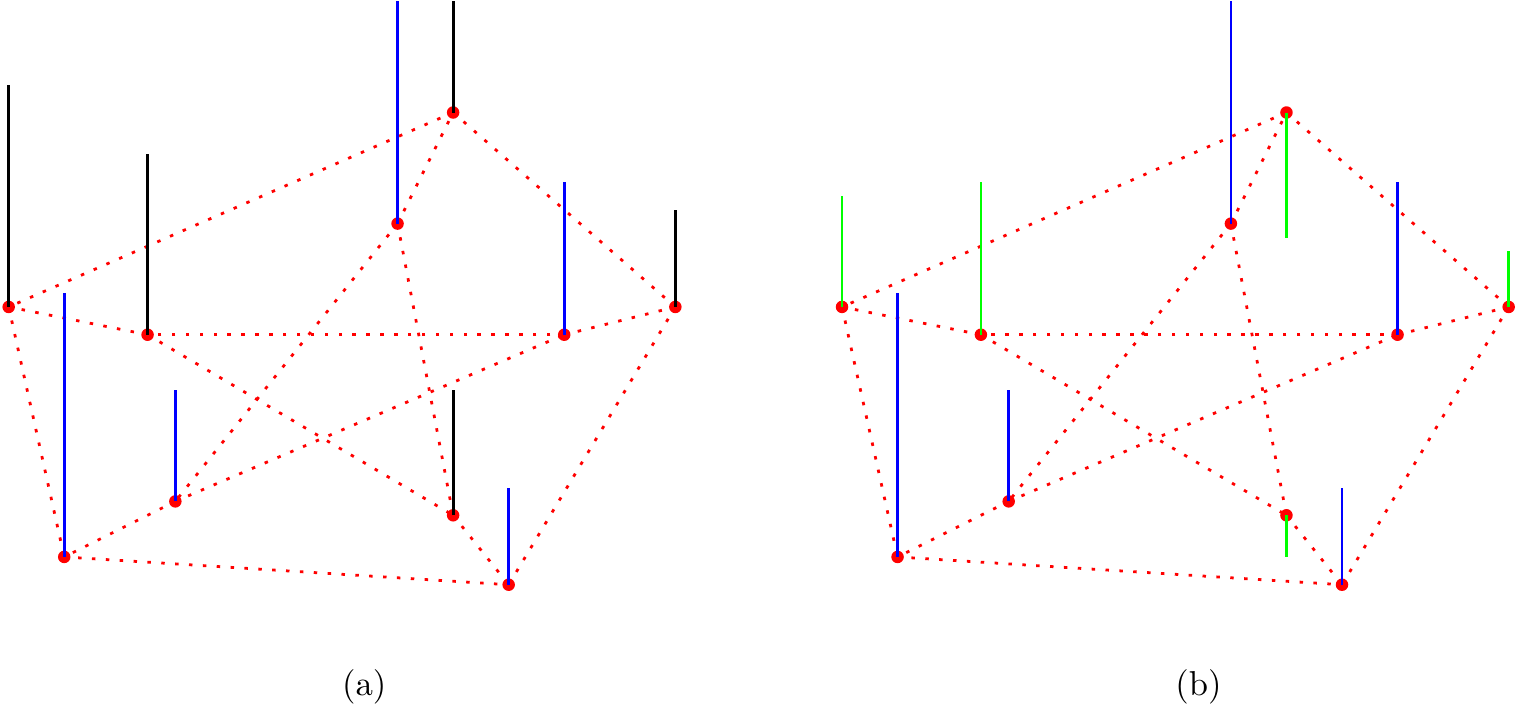}
\caption{In both images, the blue signals are observed on a subset of vertices. The extended signals include the black signals in (a) and green signals in (b). In \cref{it:lzb}, we choose (a) as the more natural extension of the observed subgraph signals.}
\label{fig:ssp17}
\end{figure}

The choice we made in \cref{it:lzb} is based on the consideration that if $i\leq j$, then $y_i$ is understood to be ``smoother'' than $y_j$ as signals on $G$. However, if $x_i$ and $x_j$ are similar in cosine distance, we need to make a choice between $y_i$ and $y_j$ as an extension. It is probably more natural to extend $x_i$ more smoothly, i.e., with $y_i$. For example, if $x_i$ is the constant signal, it is reasonable to extend it to a constant signal on $G$. An illustration is given in \figref{fig:ssp17}.

With the above constructed $\Omega$ and taking $\mu$ to be the discrete measure, we solve the following least squares problem, where $\FV{0}$ and $\scF_{V}$ are as proposed in \cref{sec:xv0} and \cref{sec:xv}, respectively. 

\begin{prob} \label{prob:th}
\begin{align}
(F^*,F^*_0) = \argmin_{F\in \scF_V, F_0 \in \scF_{V_0}} \sum_{(x,y)\in \Omega} \norm{F_0(x)-P_{V_0}\circ F(y)}_2^2.
\end{align}
The least squares sum is denoted by $s^2_{V_0}$.
\end{prob}

\begin{Lemma}\label{lem:th_convex}
\cref{prob:th} is a convex optimization problem if the polynomial coefficients in $\scF_V$ are either unconstrained or fixed.
\end{Lemma}
\begin{IEEEproof}
Recall that from \cref{sec:xv0} and \cref{sec:xv} that $\FV{0}$ and $\scF_{V}$ are convex families of transformations. For $\scF_{V}$, with the additional constraint that some polynomial coefficients are fixed, the optimization feasible set remains convex. Furthermore, since $F_0(x)-\PV{0}F(y)$ is linear in $(F,F_0)$ for each $(x,y)$ and $\norm{\cdot}_2$ is convex, \cref{prob:th} is a minimization of a convex function over a convex set, and hence convex.
\end{IEEEproof}

\begin{Example}\label{eg:lvb}
\begin{enumerate}[a)]
\item Consider an unweighted graph $G=(V,E)$. Let $y_0,\ldots,y_{|V|-1}$ be an orthonormal eigenbasis of the Laplacian $L_G$. Suppose $V_0$ is a subset of $V$ and $H_0$ is the induced subgraph. Let $\partial V_0$ be the boundary of $V_0$, i.e., $v\in \partial V_0$ if $v\in V_0$ and some neighbor of $v$ is not in $V_0$. For $v\in \partial V_0$, denote by $d(v)$ the number of its neighbors in $V\backslash V_0$. We use the sum of squares $S = \sum_{v\in \partial V_0}d(v)^2$ to measure connectedness between $V_0$ and $V\backslash V_0$.

By \cref{defn:gas} with $r=0$, $C_{V_0} = V_0$ and $D_{V_0}=1$, we claim that if \cref{prob:th} is solved with $\scF_V= \scF_{C_{V_0}, D_{V_0}}$, the coefficient of $L_G$ in $F\in\scF_{V}$ fixed to be $1$ and $\FV{0}$ consisting of graph Laplacians, then we have $s^2_{V_0} \leq 4S$. To see this, it suffices to consider the special choices $F = L_G$ and $F_0 = L_{H_0}$ and we have
\begin{align*}
s^2_{V_0} &\leq  \sum_{0\leq i\leq |V|-1} \norm{F_0\circ P_{V_0}(y_i) - P_{V_0}\circ F(y_i)}_2^2 \\
&=  \sum_{0\leq i\leq |V|-1} \sum_{v\in V_0} (F_0\circ P_{V_0}(y_i)(v) - P_{V_0}\circ F(y_i)(v))^2\\
& = \sum_{0\leq i\leq |V|-1} \sum_{v\in \partial V_0} (F_0\circ P_{V_0}(y_i)(v) - P_{V_0}\circ F(y_i)(v))^2\\
& = \sum_{0\leq i\leq |V|-1} \sum_{v\in \partial V_0} \Big(\sum_{\substack{u\in V\backslash V_0,\\ (u,v)\in E}} (y_i(v)-y_i(u))\Big)^2\\
& \leq \sum_{0\leq i\leq |V|-1} \sum_{v\in \partial V_0} d(v)\sum_{\substack{u\in V\backslash V_0,\\ (u,v)\in E}} \Big( y_i(v)-y_i(u)\Big)^2\\
& \leq \sum_{0\leq i\leq |V|-1} \sum_{v\in \partial V_0} d(v)\cdot 2\sum_{\substack{u\in V\backslash V_0,\\ (u,v)\in E}} (y_i(v)^2+y_i(u)^2) \\
& = 2\sum_{v\in \partial V_0} d(v) \sum_{\substack{u\in V\backslash V_0,\\ (u,v)\in E}}\Big( \sum_{0\leq i\leq |V|-1}y_i(v)^2 + \sum_{0\leq i\leq |V|-1}y_i(u)^2 \Big) \\ 
& = 2\sum_{v\in \partial V_0} d(v) \sum_{\substack{u\in V\backslash V_0,\\ (u,v)\in E}} 2 = 4\sum_{v\in \partial V_0}d(v)^2 = 4S.
\end{align*}
where the second equality is because $F_0\circ P_{V_0}(y_i)(v) - P_{V_0}\circ F(y_i)(v)=0$ for $v\in V_0\backslash \partial V_0$, and the second inequality follows from $(a_1+\ldots + a_m)^2\leq m(a_1^2+\ldots+a_m^2)$. Moreover, we use the fact that $\sum_{0\leq i\leq |V|-1}y_i(v)^2=1$ for each $v\in V$ in the third last equality. We see that if $S$ is small, then letting $F_0=L_{H_0}$ already gives a small $s_{V_0}^2$. 

\item Recall \cref{eg:lgbt} where $|V_0|$ divides $|V|$. To fit with the classical DFT scenario, we consider the adjacency matrix of a directed cycle graph $G$ to be the shift operator $L_G$. Following \cref{sec:xv}, we set $\scF_V = \scF_{C_{V_0},D_{V_0}}$ with $C_{V_0} = V_{|V|/|V_0|} = V_0$ as in \cref{defn:gas} and $D_{V_0} = |V|/|V_0|$. To avoid trivial solutions, we require the leading coefficient of $F \in \scF_V$, as a polynomial of $L_G$, to be $1$. For $\scF_{V_0}$, we  let it consist of the adjacency matrices of constant weight directed graphs. If we set $\delta = 0$ in \cref{it:lzb} in the construction of $\Omega$, due to periodicity, we have exactly $|V_0|$ pairs $(x_i,y_i), 0\leq i\leq |V_0|-1$, for eigenvectors $y_i$ corresponding to the lowest $|V_0|$ frequencies. Solving \cref{prob:th} with such a setup, we recover \cref{eg:lgbt} with $s^2_{V_0} = 0$. 
\end{enumerate}
\end{Example}

\subsection{Filter Learning with Data} \label{sec:fil}

In \cref{prob:thh,prob:th}, our objective is to learn a shift operator $F_{0}$ on the induced subgraph $H_0$ of $V_0$ and its corresponding filter $F$ on $G$ that it is consistent with. Here, we illustrate that our framework can also be adapted to a data-driven context. In this problem formulation, we consider learning a filter $F_0$ on $H_0$ that approximates an unknown filter $F$ on $G$ from data generated by $F$.

Suppose we have graph signals $y_t$ and $z_t$ related by $z_t = F(y_t)$, $t=1,\ldots,T$, for some unknown $F\in\scF_V$. We have access to only the partial observations $x_t=\PV{0}(y_t)$ and $x'_t=\PV{0}(z_t)$, for $t=1,\ldots,T$. Our objective is to infer the filter $F$ as well as to find a filter $F_0$ on $V_0$ to approximate $F$. We propose to solve the following optimization problem.
\begin{subequations}\label{prob0}
\begin{align}
\min_{F\in \scF_V, F_0\in \scF_{V_0}}\ & \sum_{t=1}^T \norm{x'_t - F_0(x_t)}_2^2, \label{prob0_obj}\\
\st\ & \ell(P_{V_0}\circ F, F_0\circ P_{V_0})\leq \epsilon, \label{prob0_cons}
\end{align}
\end{subequations}
where $\epsilon$ is a positive constant. The loss $\ell$ acts as a regularizer if we consider the Lagrangian form:
\begin{prob} \label{prob:filter}
\begin{align} \label{eq:mff}
\min_{F\in \scF_V, F_0\in \scF_{V_0}}\ & \sum_{t=1}^T \norm{x'_t - F_0(x_{t})}_2^2 + \beta \ell(P_{V_0}\circ F, F_0\circ P_{V_0}),
\end{align}
for some positive constant $\beta$. 
\end{prob}
\cref{prob:filter} is useful when it is impossible to recover the full graph signals $y_t$ and $z_t$ from $x_t$ and $x'_t$, respectively. On the other hand, viewing $x_t$ and $x'_t$ as graph signals independent of $y_t$ and $z_t$, and solving \cref{prob0_obj} without the constraint \cref{prob0_cons} (as is typically done in GSP), ignores the prior knowledge that $x_t,x'_t$ are generated from $y_t,z_t$.


\section{Numerical results} \label{sec:sim}

In this section, we present numerical experiments to illustrate the subGSP framework and its applications. For all the experiments, we use the following choices for $\scF_{V}$ and $\scF_{V_0}$: 
\begin{itemize}
\item $\scF_{V}$: First, we form $C_{V_0}$ and $D_{V_0}$ with $r = 2$ as in \cref{defn:gas}. Take $C$ to be the refinement of $C_{V_0}$ by including $U_i= V_i\backslash V_{i+1}$, for each $i\geq 1$. The degree associated with $U_i$ is the same as that of $V_i$. Furthermore, we constrain the coefficient of $\oPV{1}\circ L_G$ to be $1$.

\item $\scF_{V_0}$: We use choice~\ref{it:xym} of \cref{sec:xv0}. 
\end{itemize}

In most of the applications, we solve \cref{prob:th} or \cref{prob:filter} to obtain $(F^*,F^*_0)$, and use $F^*_0$ and its associated eigenbasis in our signal processing tasks. %
For comparison, we consider the same procedure applied to the operators $L_{H_0}$ and $K$ in place of $F^*_0$, where $L_{H_0}$ is the Laplacian of the induced graph $H_0$ of $V_0$, and $K$ is the Kron reduced Laplacian matrix \gls{wrt} $V_0$. 

The graphs used in the experiments include synthetic Girvan–Newman (GM) community graphs with $3$ and $4$ components of size $120$ \cite{Gir02} (denoted as 3-GM and 4-GM respectively), a square lattice graph, the Enron graph with $500$ vertices,\footnote{https://snap.stanford.edu/data/email-Enron.html} a part of the Arizona power network of size $47$ \cite{Ari12}, and a USA weather station network of size $197$.\footnote{http://www.ncdc.noaa.gov/data-access/land-based-station-data/station-metadata} The graph for the Arizona power network is based on physical connections between the power components representing the vertices of the graph; and the graph for the weather station network is constructed using the $k$-nearest neighbor algorithm based on topographic locations of the stations.

\subsection{Spectral Distribution of Subgraph Shift}\label{sec:spectral}

In this experiment, we study the spectral distributions of $F^*$ and $F^*_0$ after solving \cref{prob:th} with the prescribed $\scF_V$ and $\scF_{V_0}$. As $F^*$ models an (asymmetric) shift on $G$, we investigate how the eigenvalues of $F^*_0$ compare with the main spectral set of $F^*$ (\cref{defn:tls}). For example, by our choice of $\scF_{V_0}$, though $F^*_0$ is symmetric, it is {\it a priori} not positive semi-definite. We want to see whether most of the eigenvalues are positive or not. 

\begin{table}[!htb]
\caption{Graph statistics} \label{tab:ssp1}
\centering  
\begin{tabular}{|l|c|c|c|c|c|c|c|}  
\hline
\emph{Graphs $G$} & deg. & $p$ & no. comp. & main comp. size & avg. $|V_0|$\\ 
\hline\hline
$3$-GM graph & $5.6$ & $0.4$ & $7.7$ & $89.5\%$ & $47.8$\\
\hline
$4$-GM graph & $5.2$ & $0.4$ & $8.3$ & $88.6\%$ & $47.5$ \\
\hline
Square lattice & $3.7$ &0.4& $18.9$ & $82.9\%$ & $57.3$  \\
\hline
Enron graph & $12.6$ &0.2& $33.2$ & $69.6\%$ & $98.7$ \\
\hline
\end{tabular}
\end{table}   

We consider two types of GM community graphs, square lattice and the Enron graph. The former two graphs have clear community structures, while the latter two do not. Moreover, the Enron graph is a small world graph, in contrast to the lattice graph. The graph statistics are summarized in \cref{tab:ssp1}. In the table, ``deg.'' is the average degree of the graph $G$, $p$ is the the probability that a node of $V$ is included in $V_0$, ``no. comp.'' is the average number of connected components of $H_0$, ``main. comp. size'' is average size of the main component of $F^*$ measured as a percentage of $|V_0|$, and ``avg. $|V_0|$'' is the average size of $V_0$.

\begin{figure}[!htbp]
\centering
\begin{subfigure}[b]{.5\linewidth}
\centering
\includegraphics[width=.9\columnwidth,trim=0cm 6cm 1cm 6.5cm,clip]{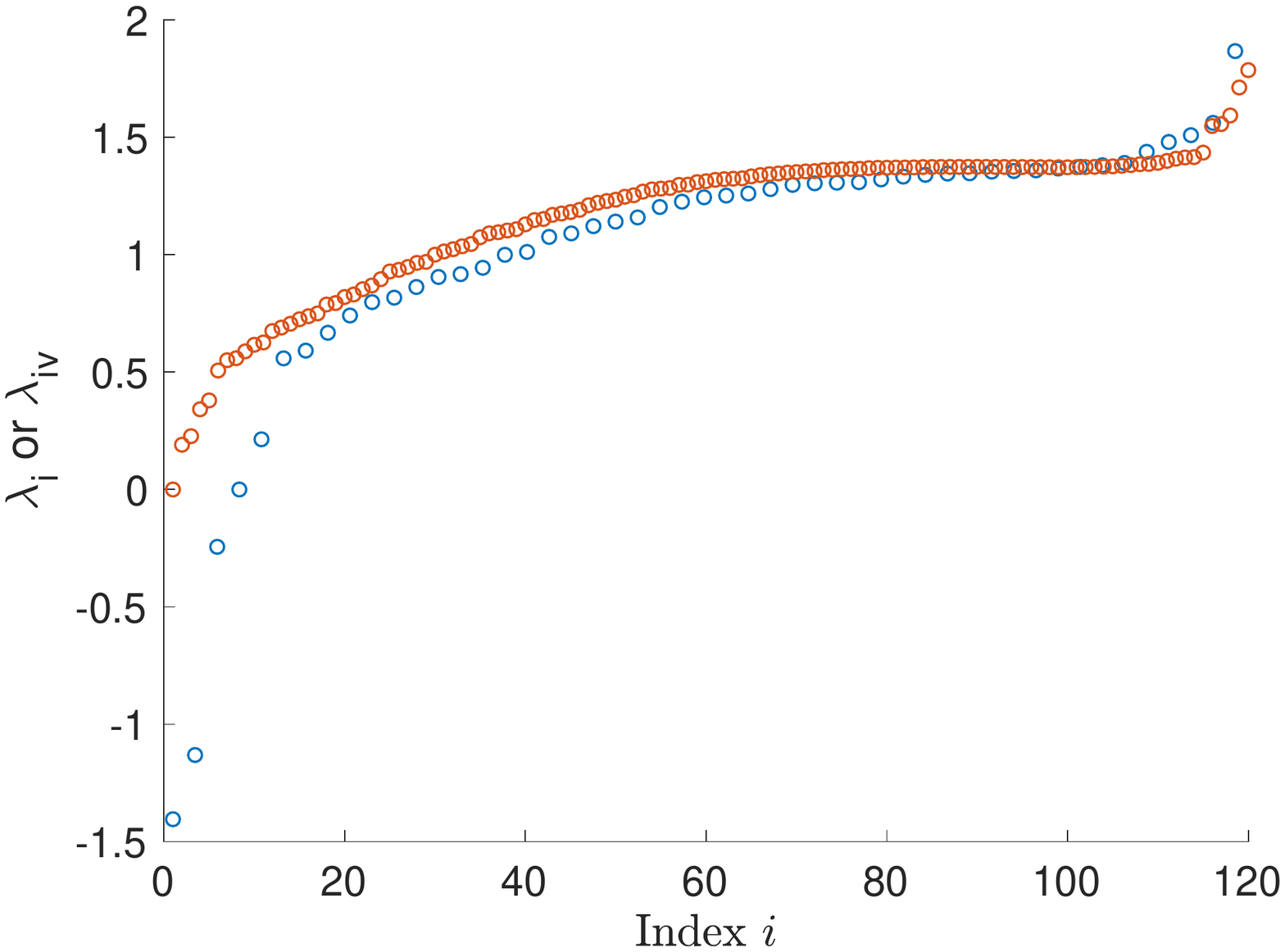}
\caption{$3$-GM graph}
\end{subfigure}%
\begin{subfigure}[b]{.5\linewidth}
\centering
\includegraphics[width=.9\columnwidth,trim=0cm 6cm 1cm 6.5cm,clip]{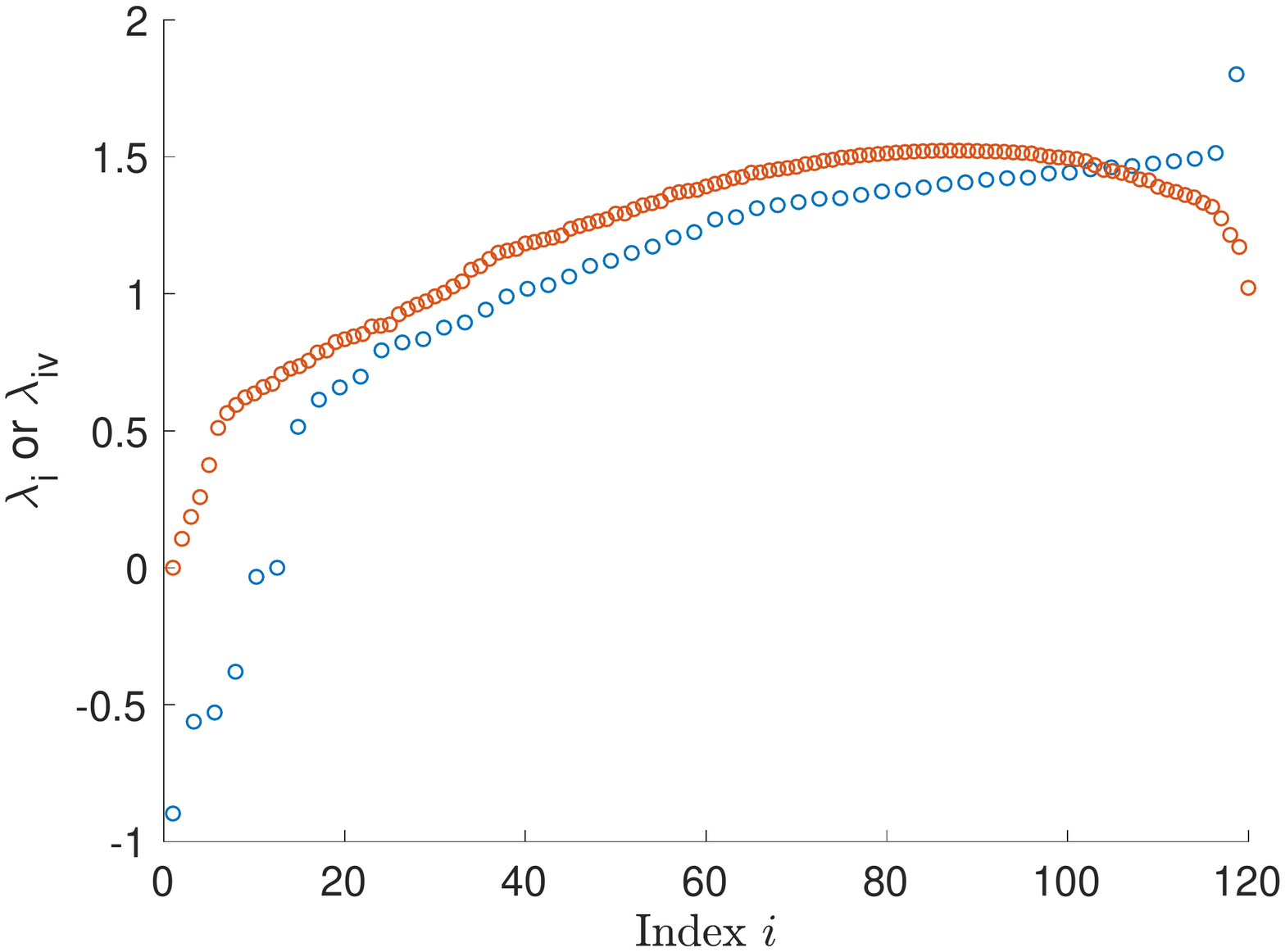}
\caption{$4$-GM graph}
\end{subfigure}%
\newline
\begin{subfigure}[b]{.5\linewidth}
\centering
\includegraphics[width=.9\columnwidth,trim=0cm 6cm 1cm 6.5cm,clip]{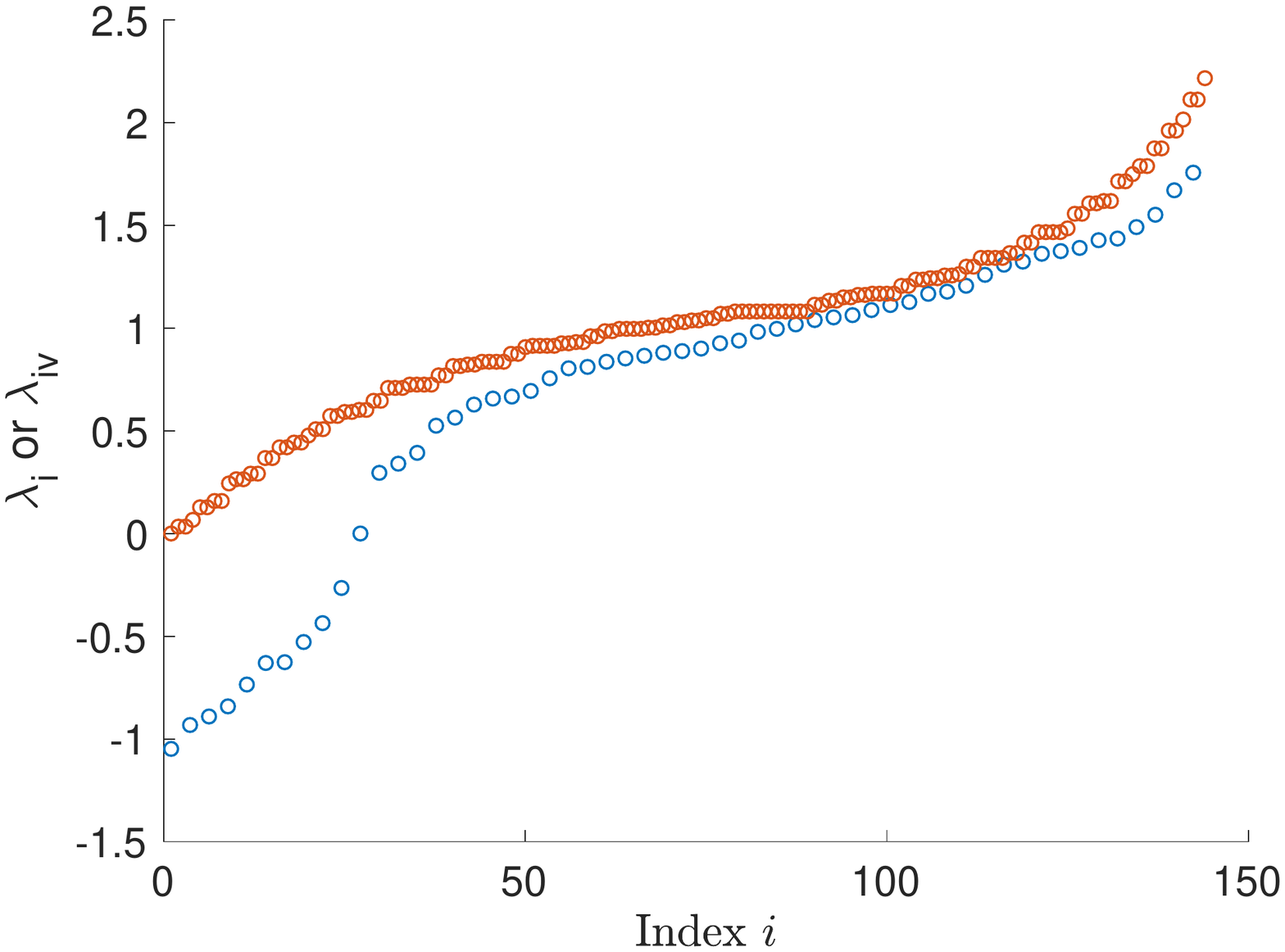}
\caption{Square lattice}		
\end{subfigure}%
\begin{subfigure}[b]{.5\linewidth}
\centering
\includegraphics[width=.9\columnwidth,trim=0cm 6cm 1cm 6.5cm,clip]{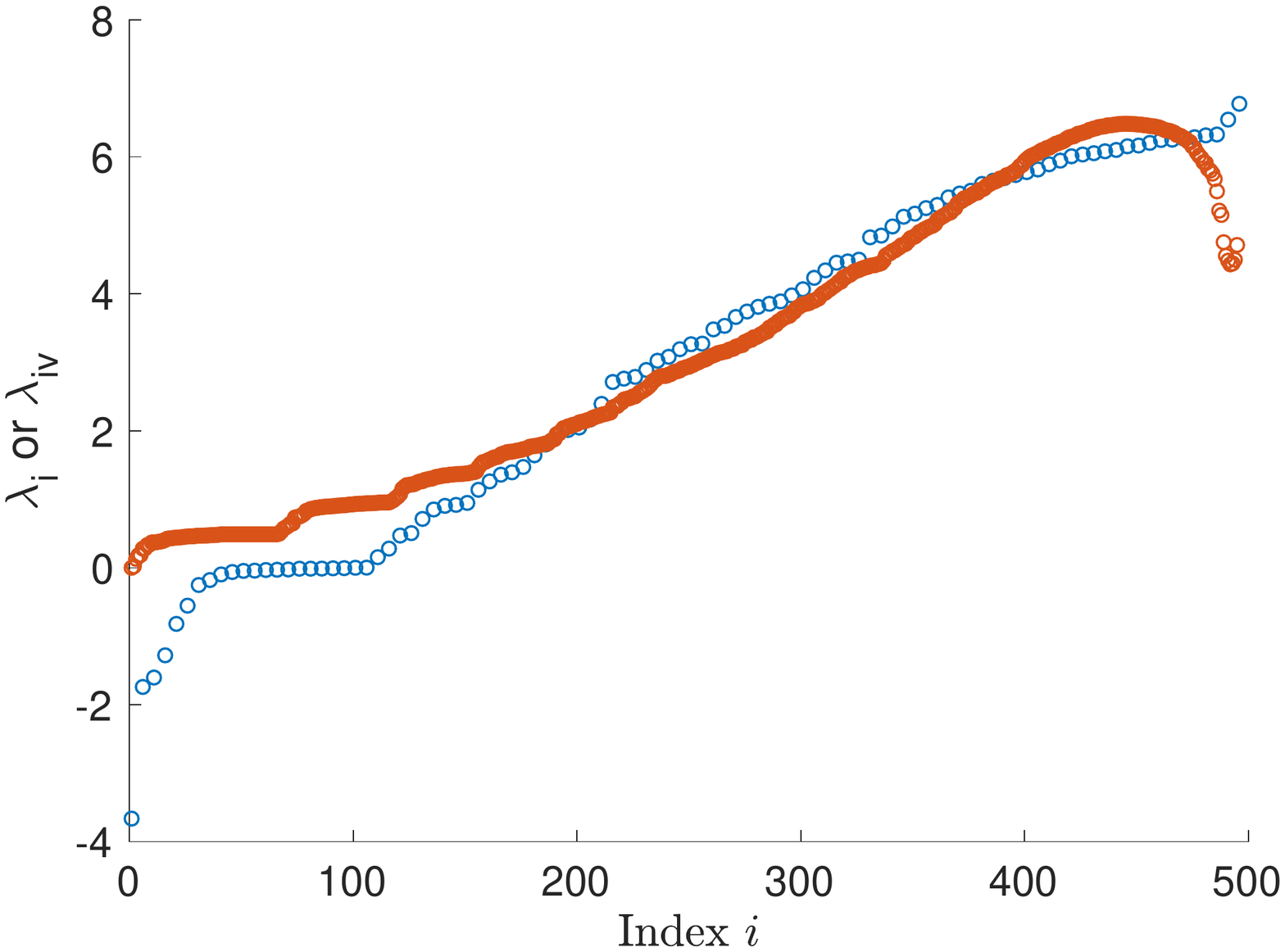}
\caption{Email graph}
\end{subfigure}%
\caption{The plot of the eigenvalues of $F^*_0$ (blue) and the main spectral set of $F^*$ (orange).}\label{fig:ssp2}
\label{fig:2}
\end{figure}

We randomly choose $V_0$, and solve \cref{prob:th} to obtain $(F^*,F^*_0)$. From \cref{tab:ssp1}, we see that the main spectral set size occupies a large proportion of $|V|$. In comparison, from \cref{fig:ssp2}, we notice that the eigenvalues of $F^*_0$ fit largely with the main spectral set of $F^*$, especially for large graph frequencies. The main spectral set of $F^*$ consists of non-negative values, and most of the eigenvalues of $F^*_0$ are non-negative as well. This indicates that $F^*_0$ still enjoys some important features of a Laplacian. 

\subsection{Signal Compression}

In this experiment, we study the problem of signal compression on a subgraph. For a graph $G=(V,E)$, suppose we have a smooth signal $y$ that is bandlimited \gls{wrt} the graph Laplacian $L_G$. Let $V_0\subset V$. We only observe $x = \PV{0}(y)$ and our objective is to compress the signal $x$.  The exact bandlimit of $y$ is assumed to be unknown \emph{a priori}. 

\begin{table}[!htb]
\caption{Signal Compression Error for $\theta_c=0.4$} \label{tab:ssp2}
\centering  
\begin{tabular}{|l|c|c|c|} 
\hline
\emph{Subgraph Shift} & $F^*_0$ & $L_{H_0}$ & $K$ \\ 
\hline\hline
$3$-GM graph & $15.6\%$ & $19.2\%$ & $61.8\%$ \\
\hline
$4$-GM graph & $6.5\%$ & $11.1\%$ & $24.3\%$  \\
\hline
Square lattice & $18.2\%$ &$27.1\%$& $34.7\%$  \\
\hline
\end{tabular}
\end{table}  

We solve \cref{prob:th} to obtain $(F^*,F^*_0)$. We then decompose $x$ w.r.t.\ an (ordered) orthonormal eigenbasis $\{x_i \mid 0\leq i\leq  |V_0|-1 \}$ of $F^*_0$ as $x = \sum_{0\leq i\leq |V_0|-1} \hat{x}(i)x_i$ (c.f.\ \cref{sec:sub}). We perform compression of $x$ by retaining only the first $\theta_c$ fraction of Fourier coefficients to obtain 
\begin{align*}
x_c = \sum_{0\leq i\leq \floor{\theta_c|V_0|}-1} \hat{x}(i)x_i.
\end{align*}
The compression error is defined to be $\norm{x-x_c}_2/\norm{x}_2$. We perform the same procedure for $L_{H_0}$ and $K$ in place of $F^*_0$ as baseline comparisons. 

\begin{figure}[!htb]
\centering
\begin{tikzpicture}[baseline]
\begin{axis}[
title={Signal Compression Error},
xlabel={$\theta_c$}, 
ylabel={error ($\%$)},
xmin=0.35, xmax=0.75,
ymin=0, ymax=30,
xtick={0.4,0.5,0.6,0.7},
ytick={0, 5, 10, 15, 20, 25, 30},
legend pos=north east,
legend style={font=\scriptsize},
ymajorgrids=true,
grid style=dashed,
]

\addplot
[
color=blue,
mark=square,
]
coordinates {
(0.4,15.6)(0.5,8.09)(0.6, 5.0)(0.7,9.2)
};
\addlegendentry{$F^*_0$, $3$-GM graph}

\addplot
[
color=blue,
mark= *,
]
coordinates {
(0.4,19.2)(0.5,10.2)(0.6, 5.7)(0.7,5.5)
};
\addlegendentry{$L_{H_o}$, $3$-GM graph}

\addplot
[
color=red,
mark=square,
]
coordinates {
(0.4,6.5)(0.5,6.5)(0.6, 8.0)(0.7,6.6)
};
\addlegendentry{$F^*_0$, $4$-GM graph}

\addplot
[
color=red,
mark=*,
]
coordinates {
(0.4,11.1)(0.5,9.6)(0.6, 7.2)(0.7,7.9)
};
\addlegendentry{$L_{H_o}$, $4$-GM graph}

\addplot
[
color=green,
mark=square,
]
coordinates {
(0.4,18.2)(0.5,10.6)(0.6, 8.9)(0.7,7.0)
};
\addlegendentry{$F^*_0$, Square lattice}

\addplot
[
color=green,
mark=*,
]
coordinates {
(0.4,27.1)(0.5,11.8)(0.6, 9.5)(0.7,7.2)
};
\addlegendentry{$L_{H_o}$, Square lattice}

\end{axis}
\end{tikzpicture}
\caption{Signal compression error for procedures based on $F^*_0$ and $L_{H_0}$ as the subgraph shift operators. The compression error based on $K$ is omitted due to its poor performance.} 
\label{fig:ssp14}
\end{figure}
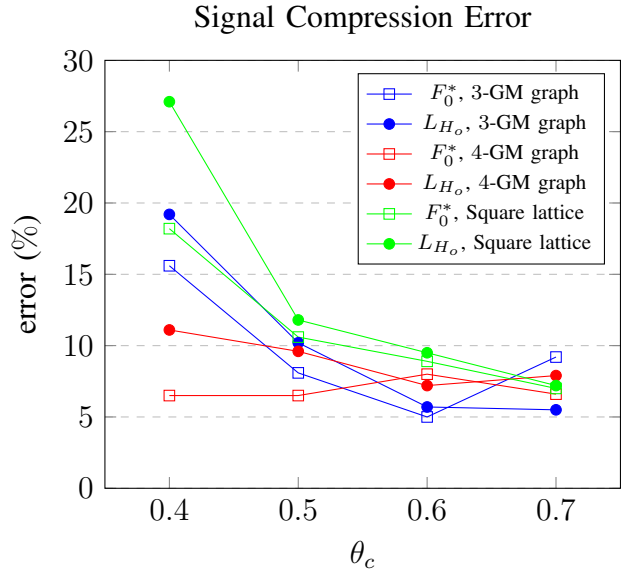

We perform simulations on GM community graphs (with $3$ and $4$ communities) and square lattice graph, with $|V_0| \approx 0.4|V|$ and $\theta_c=0.4$. We generate the nonzero GFT coefficients $\hat{y}(i)$ uniform randomly in the interval $[0,1]$ to obtain $y$ and apply the above compression procedure on $x = \PV{0}(y)$. From \cref{tab:ssp2}, we see that for each graph, the average compression error based on $F^*_0$ is the smallest. The results agree with our speculation that the smoothness of the eigenvectors of $F^*_0$ aligns with that of $L_G$.

For further investigation, we repeat the experiments with different values of $\theta_c$, with results shown in \cref{fig:ssp14}. We omit the compression based on $K$ as it has a much worse performance. As $\theta_c$ increases, there are cases where $L_{H_0}$ may be slightly better.

\subsection{Anomaly Detection} \label{sec:ano}
We now consider the task of anomaly detection with the following setting. Suppose $y$ is a graph signal on $G$, which is smooth w.r.t.\ the topology of $G$. Again, we only observe the signals at $V_0$, denoted by $x=\PV{0}(y)$. We introduce anomaly to $x$ by randomly perturbing the value of $x$ at a single vertex, i.e., changing the signal by a fixed amount $p$, and the resulting signal is denoted by $x_a$. We shall later investigate the performance for different $p$.

A classical signal processing approach for anomaly detection is to look at the high frequency components of the spectrum of $x_a$, decomposed with a suitably chosen graph shift operator. More specifically, we solve \cref{prob:th} to obtain $(F^*,F^*_0)$ and let $\{x_i \mid 0\leq i\leq |V_0|-1\}$ be an orthonormal eigenbasis consisting of eigenvectors of $F^*_0$, ordered according to their corresponding eigenvalues. The Fourier coefficients are $\{\hat{x}(i) \mid 0\leq i\leq |V_0|-1\}$ with $\hat{x}(i) = \langle x, x_i\rangle$. We choose $0<\theta_a<1$ and let $m(x) = \max_{\theta_a|V_0|\leq i\leq |V_0|-1} |\hat{x}(i)|$. With a fixed threshold $\tau>1$, we declare that $x_a$ is abnormal if $m(x_a)/m(x)>\tau$.  The exact same procedure can be applied for $L_{H_0}$ or $K$.

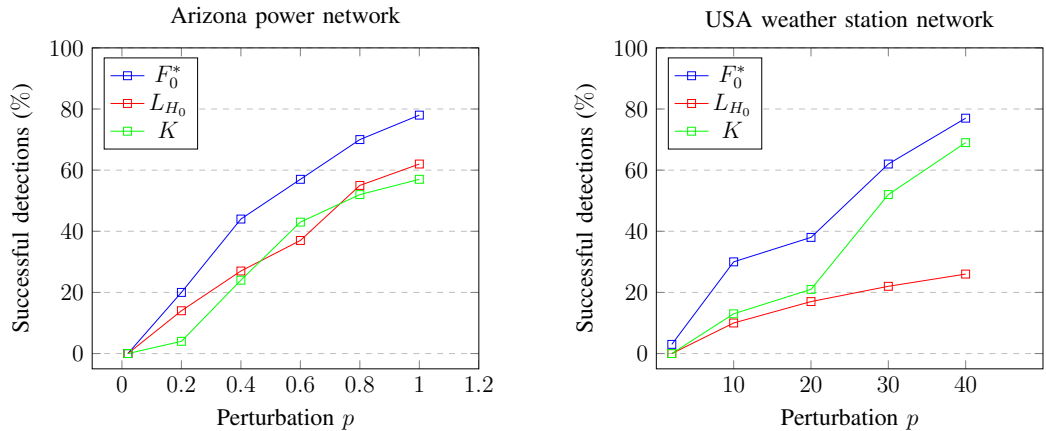
\begin{figure}[!htb]
\centering
\begin{tikzpicture}[scale = 0.75][baseline]
\begin{axis}[
title={Arizona power network},
xlabel={Perturbation $p$}, 
ylabel={Successful detections ($\%$)},
xmin=-0.1, xmax=1.2,
ymin=-5, ymax=100,
xtick={0,0.2,0.4,0.6,0.8,1.0,1.2},
ytick={0, 20, 40, 60, 80, 100},
legend pos=north west,
ymajorgrids=true,
grid style=dashed,
]

\addplot
[
color=blue,
mark=square,
]
coordinates {
(0.02,0)(0.2,20)(0.4,44)(0.6, 57)(0.8,70)(1.0,78)
};
\addlegendentry{$F^*_0$}

\addplot
[
color = red,
mark = square,
]
coordinates{
(0.02, 0)(0.2,14)(0.4,27)(0.6, 37)(0.8,55)(1.0,62)
};
\addlegendentry{$L_{H_0}$}

\addplot
[
color = green,
mark = square,
]
coordinates{
(0.02,0)(0.2,4)(0.4,24)(0.6, 43)(0.8,52)(1.0,57)
};
\addlegendentry{$K$}    

\end{axis}

\begin{scope}[xshift=10cm]
\begin{axis}[
title={USA weather station network},
xlabel={Perturbation $p$}, 
ylabel={Successful detections ($\%$)},
xmin=0, xmax=50,
ymin=-5, ymax=100,
xtick={10,20,30,40},
ytick={0, 20, 40, 60, 80, 100},
legend pos=north west,
ymajorgrids=true,
grid style=dashed,
]

\addplot
[
color=blue,
mark=square,
]
coordinates {
(2,3)(10,30)(20,38)(30, 62)(40,77)
};
\addlegendentry{$F^*_0$}

\addplot
[
color = red,
mark = square,
]
coordinates{
(2,0)(10,10)(20,17)(30, 22)(40,26)
};
\addlegendentry{$L_{H_0}$}

\addplot
[
color = green,
mark = square,
]
coordinates{
(2,0)(10,13)(20,21)(30, 52)(40,69)
};
\addlegendentry{$K$}    

\end{axis}

\end{scope}
\end{tikzpicture}
\caption{Performance of anomaly detection with $F^*_0, L_{H_0}, K$ as the subgraph shift operators.} 
\label{fig:ssp9}
\end{figure}

\begin{figure}[!htb]
\centering
\begin{subfigure}[b]{.5\columnwidth}
\centering
\includegraphics[width=.95\columnwidth, trim=0cm 6cm 0cm 7cm,clip]{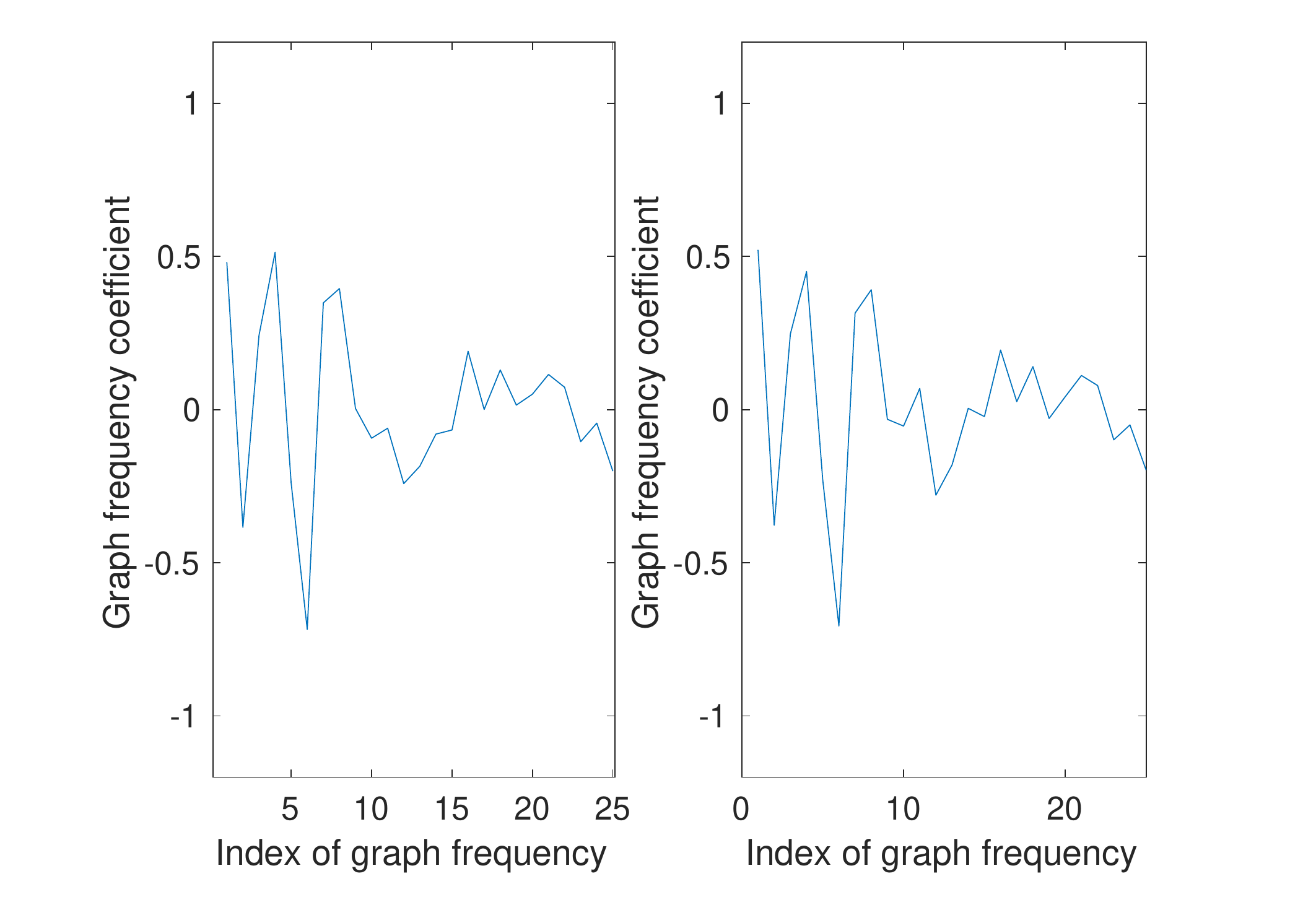}
\caption{Arizona power network}
\end{subfigure}%
\begin{subfigure}[b]{.5\columnwidth}
\centering
\includegraphics[width=.95\columnwidth, trim=0cm 6cm 0cm 7cm,clip]{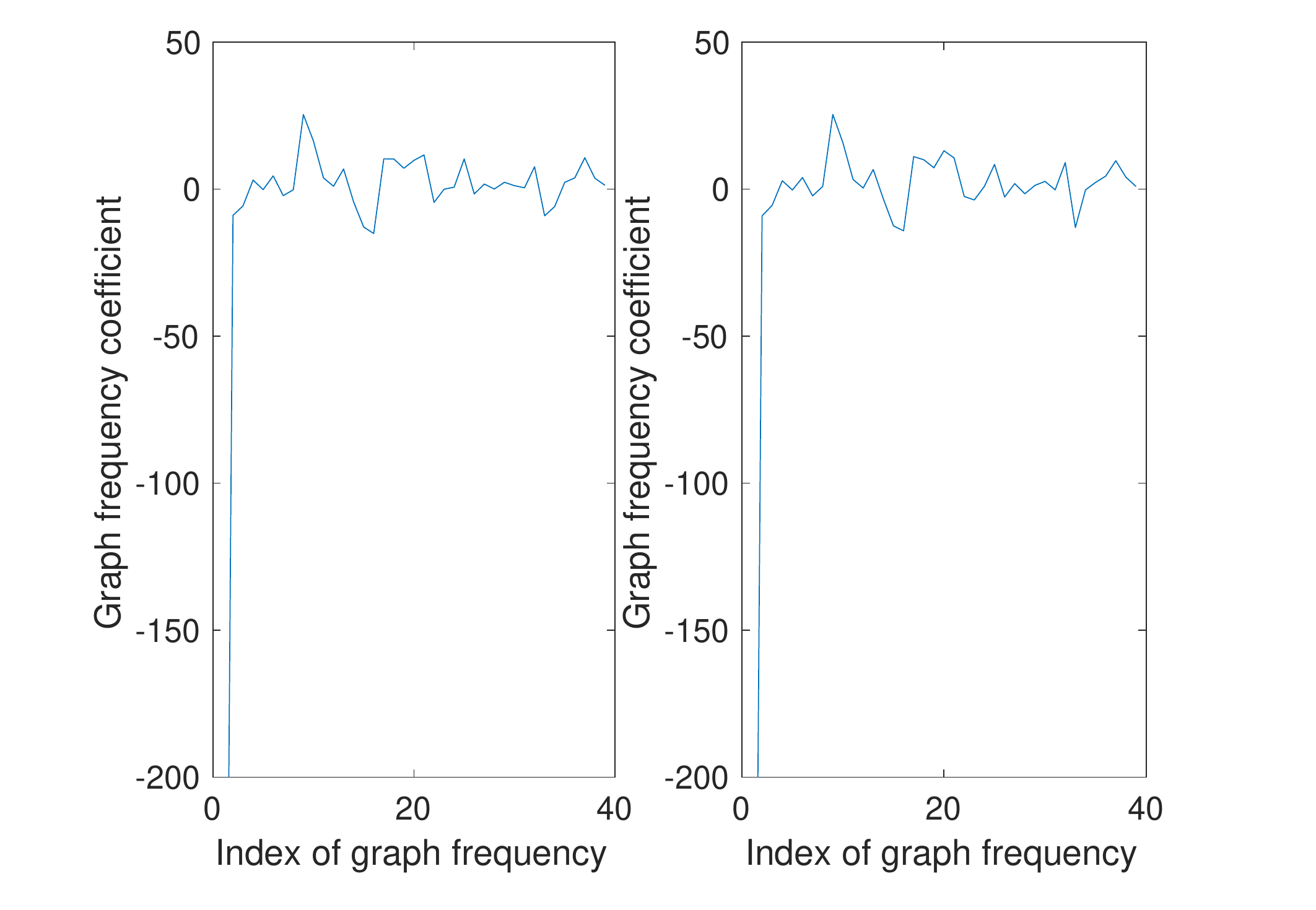}
\caption{USA weather station network}
\end{subfigure}
\caption{Examples of spectral plots of sample signals pairs $x$ (left) and $x_a$ (right) when the perturbation is small.}\label{fig:ssp8}
\end{figure} 

We run experiments on the Arizona power network $G_1$ and the USA weather station network $G_2$. The signals on $G_1$ are random bandlimited signals with small frequencies, simulating sensor recordings such as temperature. On the other hand, the signals on $G_2$ are daily temperatures recorded over the year 2013.\footnote{ftp://ftp.ncdc.noaa.gov/pub/data/gsod} The experiment parameters are set as $\theta_a=0.35, \tau=1.1, |V_0|\approx 0.5|V|$ for $G_1$; and $\theta_a=0.15, \tau=1.02, |V_0|\approx 0.2|V|$ for $G_2$. We run $200$ experiments for each graph and each $p$, and compute the percentage of successful anomaly detections. In addition, we also consider $p=0.02$ for $G_1$ and $2$ for $G_2$ to investigate false positive rate, for which all methods are low.

From \cref{fig:ssp9}, we see that the same scheme performs significantly better if we use $F^*_0$. However, the percentage of successful detection is low if the perturbation is small, e.g., $0.2$ for $G_1$ and $10$ for $G_2$. We investigate by plotting the $F^*_0$ Fourier coefficients of $x$ and $x_a$ (a typical example is shown in \cref{fig:ssp8}). We see that when the perturbation is small, the spectrum of $x$ and $x_a$ look alike. In this case, without knowing how $x_a$ is constructed, it is not possible to say it is abnormal.  

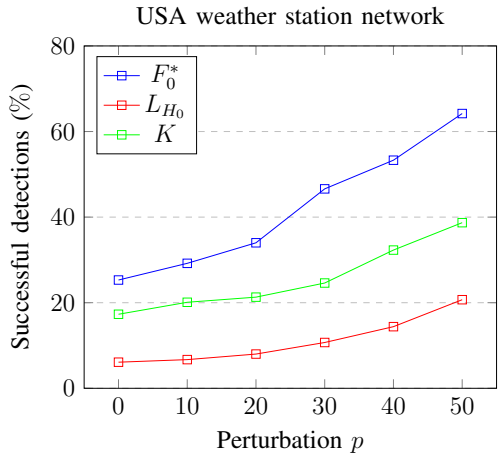
\begin{figure}[!htb]
\centering
\begin{tikzpicture}[scale = 0.8][baseline]

\begin{axis}[
title={USA weather station network},
xlabel={Perturbation $p$}, 
ylabel={Successful detections ($\%$)},
xmin=-5, xmax=55,
ymin=0, ymax=80,
xtick={0,10,20,30,40,50},
ytick={0, 20, 40, 60, 80},
legend pos=north west,
ymajorgrids=true,
grid style=dashed,
]

\addplot
[
color=blue,
mark=square,
]
coordinates {
(0, 25.3)(10, 29.2)(20, 34.0)(30, 46.6)(40, 53.3)(50, 64.2)
};
\addlegendentry{$F^*_0$}

\addplot
[
color = red,
mark = square,
]
coordinates{
(0, 6.1)(10, 6.7)(20, 8.0)(30, 10.7)(40, 14.4)(50, 20.7)
};
\addlegendentry{$L_{H_0}$}

\addplot
[
color = green,
mark = square,
]
coordinates{
(0, 17.3)(10, 20.1)(20, 21.3)(30, 24.6)(40, 32.3)(50, 38.7)
};
\addlegendentry{$K$}    

\end{axis}

\end{tikzpicture}
\caption{Performance of anomaly detection using pairs $x_{t-1}, x_{t,a}$.} 
\label{fig:anomaly}
\end{figure}  

For the weather station dataset, we may consider the following more realistic approach. We consider partial readings on $V_0$ on two consecutive days $x_{t-1}, x_t$. We apply perturbation $p$ to $x_t$ to obtain $x_{t,a}$. An anomaly is declared if $m(x_{t,a})/m(x_{t-1})>\tau$ for a prescribed threshold $\tau>1$. We run experiments for randomly chosen $t$ with $\theta_a=0.15, \tau=1.1, |V_0|\approx 0.2|V|$. The results are shown in \cref{fig:anomaly}. We see that $F^*_0$ performs significantly better for anomaly detection; however, it has a higher false positive rate at $p=0$. A greater improvement performance is seen for $F_0^*$ as $p$ increases. 

\subsection{Denoising} 

Next, we consider denoising. In contrast to anomaly detection in \cref{sec:ano}, we add white noise to the signal $x$ for every vertex of $V_0$ to form $x_{\alpha}$. For denoising, we observe $x_{\alpha}$ and want to recover $x$ as far as possible.  

As briefly described in \cref{sec:sub}, a denoising approach is to apply a convolution filter to $x_b$ in the frequency domain that scales down the high frequency components. More specifically, we first choose $0<\theta_d<1$ and a scaling factor $0\leq s_d<1$. Let $\hat{x}_{\alpha}$ be the GFT of $x_{\alpha}$ using $F^*_0$ from solving \cref{prob:th} as the shift operator. The recovered signal $\tilde{x}$ on $V_0$ is the unique signal whose $i$-th Fourier coefficient is $s_d\hat{x}_{\alpha}(i)$ for $\floor{\theta_d|V_0|}\leq i\leq |V_0|-1$ and $\hat{x}_{\alpha}(i)$ for $i< \floor{\theta_d|V_0|}$. 

To evaluate the performance, we compute the ratio between the errors of the recovered signal and noisy signal $r_e = \norm{x-\tilde{x}}_2/\norm{x-x_{\alpha}}_2$. An $r_e>1$ indicates that the denoising introduces more errors and makes matters worse. The same procedure can be applied to $L_{H_0}$ and $K$.

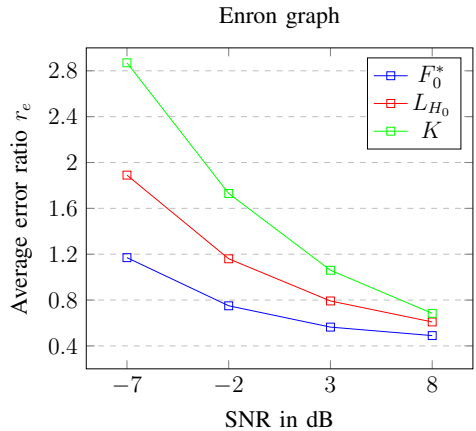
\begin{figure}[!htbp]
\centering
\begin{tikzpicture}[scale = 0.75][baseline]
\begin{axis}[
title={Enron graph},
xlabel={SNR in dB}, 
ylabel={Average error ratio $r_e$},
xmin=-9, xmax=10,
ymin=0.2, ymax=3,
xtick={-7, -2, 3, 8},
ytick={0.4, 0.8, 1.2, 1.6, 2.0, 2.4, 2.8},
legend pos=north east,
ymajorgrids=true,
grid style=dashed,
]

\addplot
[
color=blue,
mark=square,
]
coordinates {
(-7,1.17)(-2,0.75)(3, 0.564)(8, 0.490)
};
\addlegendentry{$F^*_0$}

\addplot
[
color = red,
mark = square,
]
coordinates{
(-7,1.89)(-2,1.16)(3, 0.792)(8, 0.609)
};
\addlegendentry{$L_{H_0}$}

\addplot
[
color = green,
mark = square,
]
coordinates{
(-7,2.87)(-2,1.73)(3, 1.06)(8, 0.684)
};
\addlegendentry{$K$}    

\end{axis}


%
%
%
%
\end{tikzpicture}
\caption{Performance of denoising with $F^*_0, L_{H_0}, K$ as the graph shift operators.} 
\label{fig:ssp10}
\end{figure}


We perform experiments ($100$ for each set of parameters) on the Enron graph. We consider the synthetic signals simulating timestamps of information propagation on the graph under the SI model with a random source \cite{LuoTayLen14, LuoTayLeng13, Tan18}. 
The experiment parameters are set as $\theta_d =0.2, s_d = 0.3, |V_0|\approx 0.2|V|$. %
From \cref{fig:ssp10}, we see that using $F^*_0$ achieves the best performance. However, we observe that when a relatively larger noise is introduced, the error ratio $r_e$ can be larger than $1$ even in the best case using $F^*_0$, meaning the proposed recovery increases the corruption in the signal. 

\subsection{Filter Learning}

We now consider the \cref{prob:filter} of learning a filter from partial observations. On a given graph $G=(V,E)$, we generate a shift invariant filter $\tilde{F}$ in the form $\tilde{F} = a_0+a_1L_G+a_2L_G^2$, where $L_G$ is the graph Laplacian of $G$ and $a_0, a_1, a_2$ are random coefficients chosen uniformly in the interval $[0,1]$. To generate a graph signal $y_t$, for $1\leq t \leq T$, we generate its GFT coefficients uniformly and randomly in the interval $[0,1]$. Let $z_t= \tilde{F}(y_t)$. We assume that we observe only $y_t$ and $z_t$ at a subset of vertices $V_0$. We solve \cref{prob:filter} to obtain the solution $(F^*,F_0^*)$ by setting $\ell$ as the $L^2$-loss as in \cref{prob:th}. 

\begin{figure}[!htb]
    \begin{subfigure}[b]{.5\linewidth}
    \centering
    \includegraphics[width=0.95\columnwidth,trim=0cm 6cm 0cm 6cm,clip]{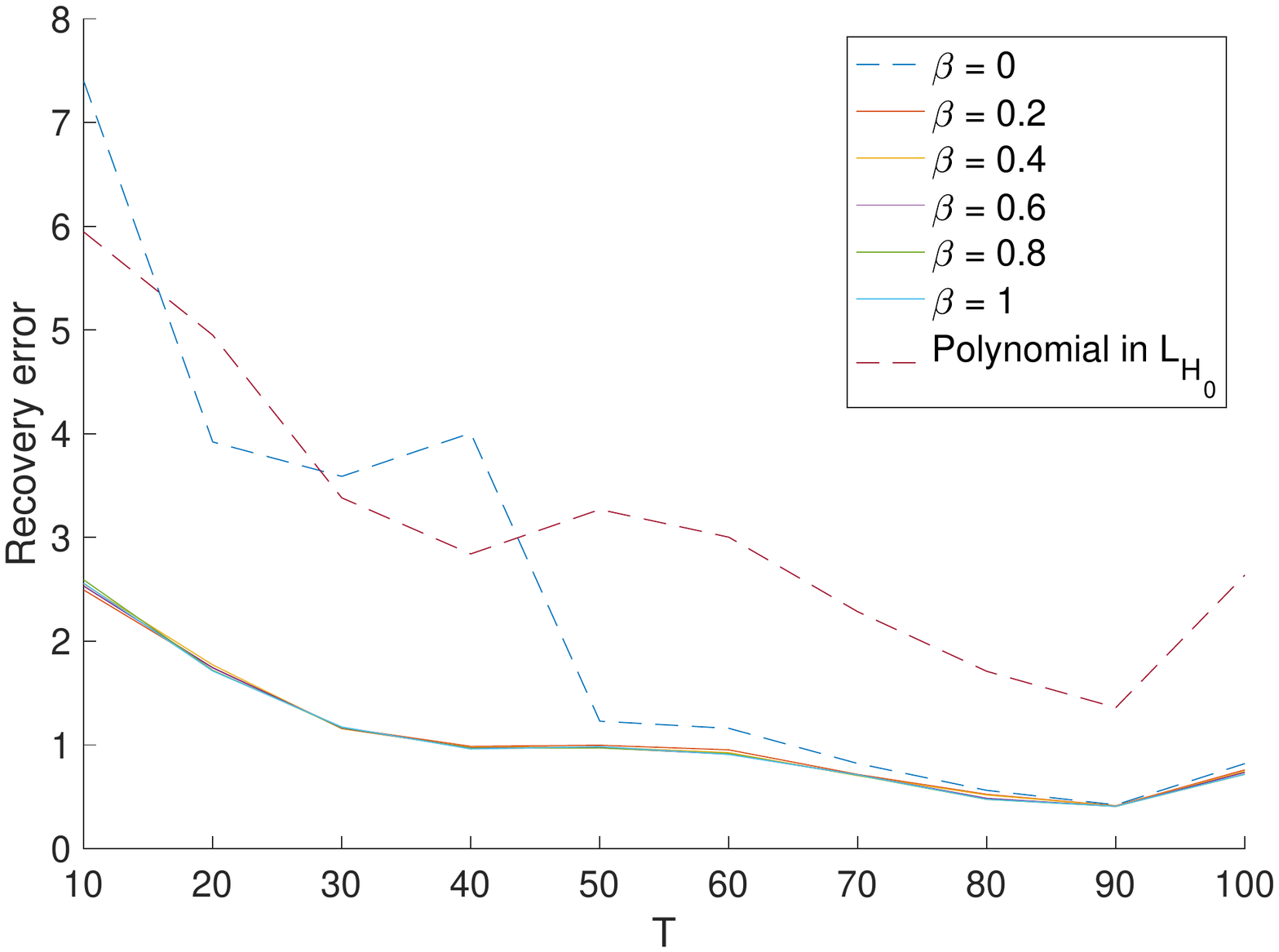}
    \caption{Square lattice}
    \end{subfigure}
    \begin{subfigure}[b]{.5\linewidth}
    \centering
    \includegraphics[width=0.95\columnwidth,trim=0cm 6cm 0cm 6cm,clip]{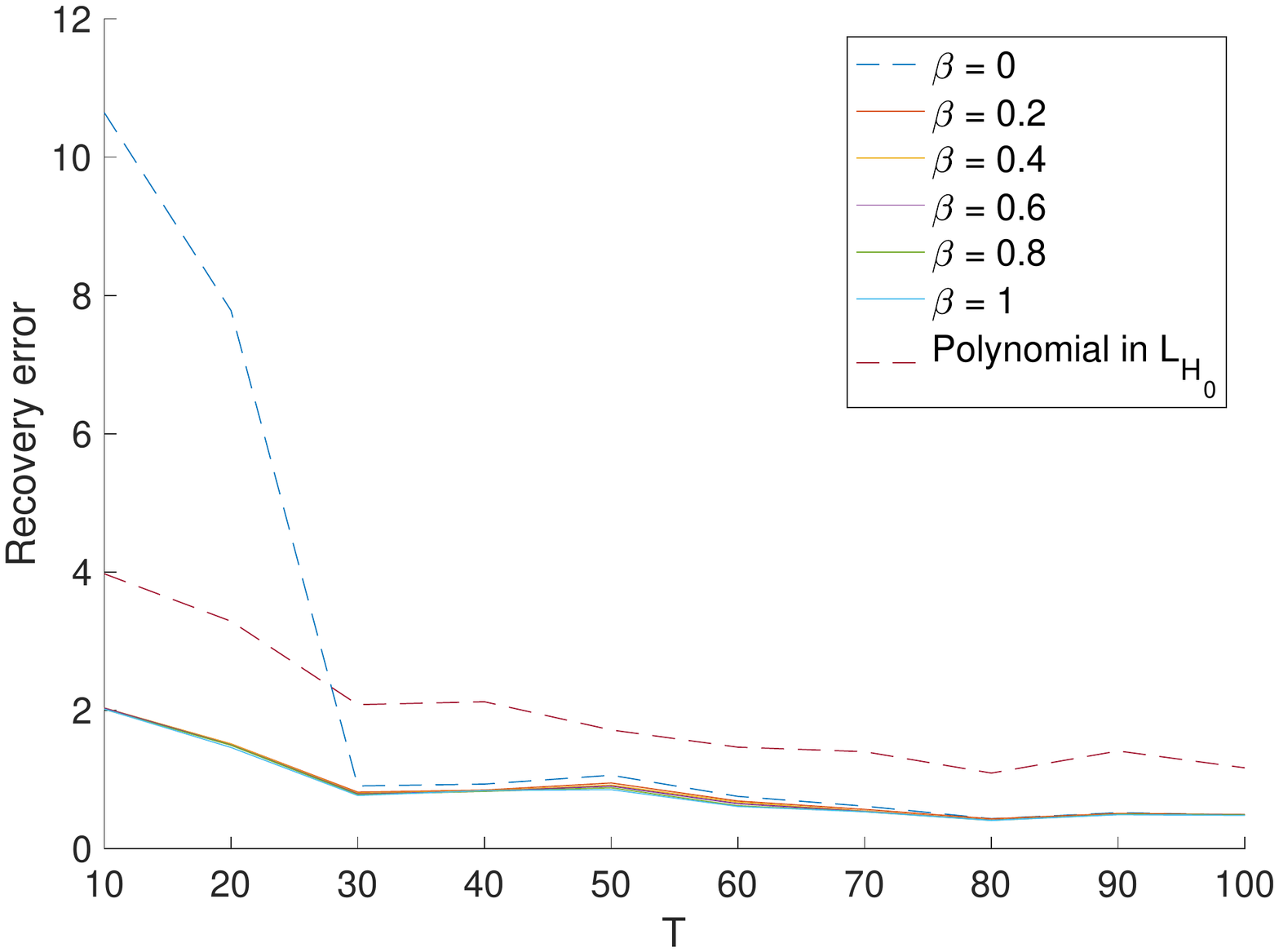}
    \caption{Power plant network}
    \end{subfigure}
    \caption{Performance of filtering learning. We highlighted the case $\beta=0$ in dashed blue curve and polynomial filter in $L_{H_0}$ in dashed red curve.}
    \label{fig:ssp15}
\end{figure}

We run $100$ experiments for each set of parameters on the square lattice and the Arizona power network respectively. The performance is evaluated by computing the average recovery error $\norm{P_{V_0}(z_t)-{F_0^*}(P_{V_0}(y_t))}_2$. For the square lattice, we set $|V_0|\approx 0.4|V|$; and for the power plant network, we set $|V_0|\approx 0.6|V|$. We test various $\beta = 0, 0.2, 0.4, 0.6, 0.8, 1$ in \cref{prob:filter} to investigate the contribution of the regularizer given by the loss $\ell$. The case $\beta=0$ corresponds to the case where there is no regularization. We also vary $T$ from $10$ to $100$. As a baseline comparison, we also learn $F_0'$ in the form of a polynomial in $L_{H_0}$. 

From \cref{fig:ssp15}, we see that by solving \cref{prob:filter} with sufficient regularization, i.e., $\beta\geq 0.2$, the average recovery error is generally smaller. The effect of regularization is more prominent when $T$ is small. Furthermore, different $\beta\geq 0.2$ have almost identical performance. We observe that our approach gives better performance than attempting to learn the filter using $L_{H_0}$. 

\section{Conclusion} \label{sec:con}

In this paper, we have proposed a subGSP framework for signals on a subset of vertices of a graph. The essential idea is to find an appropriate shift operator on the given subset. We test the performance of our approach on a few signal processing tasks with both synthetic and real graph and datasets. The results demonstrate the effectiveness of the proposed method. For future work, we want to explore further applications of our method and develop machine learning models analogous to graph neural network.


\bibliographystyle{IEEEtran}
\bibliography{IEEEabrv,StringDefinitions,allref}

\clearpage

\begin{center}
\huge \textbf{Supplementary Material}
\end{center}

\beginsupplement

\section{Random subgraph models compared} \label[Supplement]{app:con}

Suppose $G=(V,E)$ is a graph of size $n$. In this appendix, we shall compare the random vertex selection model and the Erd\H{o}s-Renyi (ER) random edge model, as initiated at the beginning of \cref{sec:sga} when we give motivation to the framework of the paper. We want to discuss when the ER model tends to produce larger components than the random vertex model, as the former is well-studied \cite{Fri04, Fan09}. 

\begin{Definition}
Let $H_0$ be an induced subgraph on a subset of vertices $V_0$ and $k$ be a positive integer. Introduce $\delta(G,H_0,k)$ to be the smallest size of an edge set $E'$ such that $G\backslash E'$ does not contain any connected subgraph of size at least $k$, other than $H_0$. Taking maximum over all connected $H_0$ of size $k$, we define 
\begin{align*}
\delta(G,k) = \max_{H_0: \text{connected of size } k} \delta(G,H_0,k).
\end{align*}   
\end{Definition}

\begin{figure}[!htb]
\centering
\includegraphics[scale=1]{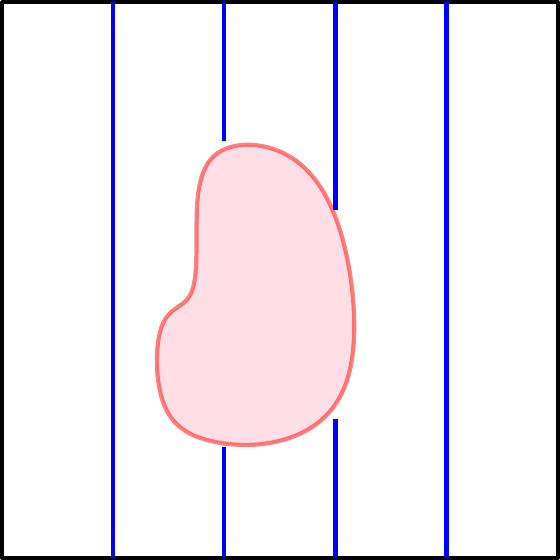}
\caption{In the illustration, the graph $H_0$ occupies the central red region and its boundary is depicted by the red curve. The boundaries of the vertical subgraphs in the argument are depicted by vertical blue lines.}
\label{fig:ssp4}
\end{figure}

\begin{Example} \label{eg:lgb}
Let $G$ be the $n\times n$ grid. Then for each $k = \calO(n^{\alpha}),\alpha<2$, we claim that
\begin{align*}
\delta(G,k) = \calO(\max(n^{3-\alpha},n^{\alpha})).
\end{align*}
To see this, let $H_0$ be a connected subgraph of order $\calO(n^{\alpha})$. Its boundary $\partial H_0$ (edges connected to the rest of the graph) is of size $\calO(n^{\alpha})$, as each vertex has degree at most $4$. On the other hand, we can cut $G$ into $\calO(n^{2-\alpha})$ vertical pieces such the size of each piece is smaller than $n^{\alpha}$. Moreover, the boundary between two adjacent vertical subgraphs has size $n$. Hence the union $E'$ of $\partial H_0$ and all the vertical boundaries has size $\calO(\max(n^{3-\alpha},n^{\alpha}))$. From the construction, it is easy to see that the size each component of $G\backslash E'$ is smaller than $n^{\alpha}$ except for $H_0$, and the claim follows. An illustration is shown in \cref{fig:ssp4}.
\end{Example}

\begin{Definition}
Given $V_0$ such that the induced subgraph $H_0$ is connected, let $\theta(G,H_0)$ be the number of distinct spanning trees of $H_0$. For any positive integer $k$, define 
\begin{align*}
\theta(G,k) = \min_{|H_0|=k} \theta(G,H_0).
\end{align*}
\end{Definition}

The number $\theta(G,k)$ can be thought of as a complexity measure of induced subgraphs of $G$ containing $k$ vertices. As two extreme cases: for every $k$, if $G$ is a tree, then $\theta(G,k)=1$; while on the other hand, if $G$ is the complete graph, then $\theta(G,k) = k^{k-2}$ by Cayley's formula. 

For $0<q<1$, denote by $G_q$ the model on induced subgraph of $G$ that includes each vertex independently with probability $q$. As in \cref{sec:sga}, let $G(q)$ be the ER model that edges are preserved independently with probability $q$. For a subgraph $H$ of $G$, let $C(H)$ be the size of the largest component of $H$. Now, we are ready to state and prove the main result. 

\begin{Theorem} \label{thm:fqa}
For $0<q_1,q_2<1$ and positive integer $k_1\leq k_2$, if 
\begin{align*}
(1-(1-q_1^{k_1-1})^{\theta(G,k_1)})q_1^{\delta(G,k_1)} \geq q_2^{k_2},
\end{align*} then 
\begin{align*}
\mathbb{P}(C(G(q_1))= k_1) \geq \mathbb{P}(C(G_{q_2})= k_2).
\end{align*}
\end{Theorem}

\begin{IEEEproof}
Let $A_1$ be the event that $C(G(q_1))= k_1$ and $A_2$ be the event that $C(G_{q_2})= k_2$. To show $\mathbb{P}(A_1) \geq \mathbb{P}(A_2)$, we introduce an ``intermediate'' event $A_3:$ 
\begin{enumerate}[(a)]
\item the largest component $C$ of $G(q_1)$ has $k_1$ vertices, and 
\item $\delta(G,k)$ edges are not in $G(q_1)$ such that the size of each other component in $G(q_1)$ other than $C$ is smaller than $k_1$.
\end{enumerate}
It is clear that $\mathbb{P}(A_1) \geq \mathbb{P}(A_3)$, and we want to show $\mathbb{P}(A_3)\geq \mathbb{P}(A_2)$.

For any $k$, let $\mathcal{C}_k$ be the sets of $k$ vertices each inducing a connected subgraph of $G$. Therefore, we can decompose $A_2$ as a union of events $A_2\subset \cup_{C\in \mathcal{C}_{k_2}}\{C\subset G_{q_2}\}$. Hence by the union bound, we have the estimate:
\begin{align*}
\mathbb{P}(A_2) & \leq \mathbb{P}(\cup_{C\in \mathcal{C}_{k_2}}\{C\subset G_{q_2}\}) \\ & \leq \sum_{C\in \mathcal{C}_{k_2}} \mathbb{P}(C\subset G_{q_2}) \\ & = \sum_{C\in \mathcal{C}_k }q_2^{k_2}.
\end{align*}

On the other hand, the event $A_3$ contains the union of events: 
\begin{align*}
\cup_{C\in \mathcal{C}_{k_1}} A_{3,C}:=\cup_{C\in \mathcal{C}_{k_1}}A_3\cap \{C\subset G(q_1) \text{ that spans a connected subgraph}\}.
\end{align*}
We observe the union is disjoint. This is because for any event in $A_3$, there cannot be two connected components of size $k_1$. In $G$, there are at least $\theta(G,k_1)$ distinct trees, which span each $C \in \mathcal{C}_{k_1}$. Therefore, the probability of each single instance in $A_{3,C}$ is at least 
\begin{align*}
(1-(1-q_1^{k_1-1})^{\theta(G,k_1)})q_1^{\delta(G,k_1)}.
\end{align*} 
Moreover, as $k_1\leq k_2$, we have $|\mathcal{C}_{k_1}|\geq |\mathcal{C}_{k_2}|$.

We now package everything together, if $	(1-(1-q_1^{k_1-1})^{\theta(G,k_1)})q_1^{\delta(G,k_1)} \geq q_2^{k_2}$, then
\begin{align*}
\mathbb{P}(A_3) & \geq \mathbb{P}(\cup_{C\in \mathcal{C}_{k_1}} A_{3,C}) = \sum_{C\in \mathcal{C}_{k_1}}\mathbb{P}( A_{3,C}) \\
& \geq \sum_{C\in \mathcal{C}_{k_1}}	(1-(1-q_1^{k_1-1})^{\theta(G,k_1)})q_1^{\delta(G,k_1)}\\
& \geq \sum_{C\in \mathcal{C}_{k_2}} q_2^{k_2} \geq \mathbb{P}(A_2).
\end{align*}
\end{IEEEproof}

\begin{Example}
As a continuation of \cref{eg:lgb}, if $k_i = \calO(n^{\alpha_i})$ for $\alpha_2 >\alpha_1 > 1.5$, then $\delta(G,k_1) = \calO(\max\{n^{3-\alpha_1},n^{\alpha_1}\}) = \calO(n^{\alpha_1})$ by \cref{eg:lgb}. Suppose we use the trivial estimation $\theta(G,k)\geq 1$. Then for any $0<q_1, q_2<1$, the condition of \cref{thm:fqa} becomes $q_1^{k_1-1}q_1^{\delta(G,k_1)}\geq q_2^{k_2}$, which is true as $n\to \infty$. Therefore, asymptotically, we always have that: $G(q_1)$ has a larger chance to contain a component of size $k_1$ than that of $G_{q_2}$ to contain a component of size $k_2$.
\end{Example}

\section{Sparsification of \texorpdfstring{$\scF_{V_0}$}{FV0}} \label[Supplement]{app:spa}

We first give the proof of \cref{prop:gva}. 
\begin{IEEEproof}[Proof of \cref{prop:gva}]
The first inequality in \cref{prop:ineq} is clear, as $\beta_{V_0,N}$ is found by minimizing over a smaller space. For the second inequality in \cref{prop:ineq}, we make use of results from the theory of spectral graph sparsification \cite{Lee15}. 

We first consider choice~\ref{it:xtl}. Recall from \cref{def:H0} that $H_0$ is the induced subgraph of $V_0$. As we are using choice~\ref{it:xtl}, $F_0^*$ is the Laplacian of a (weighted) graph $\tilde{H}$. Moreover, $\tilde{H}$ agrees with $H_0$ on $V_0$. Decompose $F_0^* = L_{\tilde{H}} = L_{H_1} + L_{H_0}$, where $L_{H_1}$ is the Laplacian of $H_1$, whose edges are those not in $H_0$. We may apply spectral matrix sparsification to $H_1$ yielding an graph $H_2$ with $N = \calO(|V_0|/\epsilon^2)$ edges whose Laplacian $L_{H_2}$ is an $\epsilon$-approximation of $L_{H_1}$, i.e., $(1-\epsilon)L_{H_1} \preceq L_{H_2} \preceq (1+\epsilon)L_{H_1}$, where $A \preceq B$ means $B-A$ is positive semi-definite. Let $H_3$ be the union $H_2\cup H_0$. In summary, (i) $L_{H_3} \in \scF_{V_0,N}$, and (ii) the sum $L_{H_3} = L_{H_2}+L_{H_0}$ is an $\epsilon$-approximation of $F_0^*$, as $L_{H_0}$ is clearly an $\epsilon$-approximation of itself. Denote $L_{H_3}$ by $F_0'$. 

For the choice~\ref{it:xtlo}, $F_0^* = L_H$ for some graph on $V_0$, we can may apply sparsification to $H$ to obtain an $\epsilon$-approximation of $F_0'$, also as the Laplacian of a sparser graph. Similarly, for the choice~\ref{it:xym}, $F_0^*$ can be written as the difference of two Laplacians $L_{H_1}-L_{H_2}$. We may apply sparsification to $H_1$ and $H_2$ to get their respective $\epsilon/2$-approximations. The difference of them, denoted by $F_0'$ is an $\epsilon$-approximation of $F_0^*$.

We can now perform the following estimation: 
\begin{align*}
\beta_{V_0,N}-\beta_{V_0} &\leq \norm{F_0'\circ P_{V_0}-P_{V_0}\circ F^*} - \norm{F_0^*\circ P_{V_0}-P_{V_0}\circ F^*} \\ 
& \leq \norm{F_0'\circ P_{V_0}-F_0^*\circ P_{V_0}} \leq \norm{F_0'-F_0^*}\\ & \leq \epsilon \norm{F_0^*} = \epsilon \mu_{\max},
\end{align*}
where $\mu_{\max}$ is largest eigenvalue of $L_{\tilde{H}}$. 

Finally, the inequality $\mu_{\max} \leq r_{\max} = 2d_{\max}$ follows directly from the definition of $\mu_{\max}$. 
\end{IEEEproof}

For the error tolerance $\epsilon r_{\max}$, the proposition allows us to perform searching in a possibly smaller subset $\scF_{V_0,N} \subset \scF_{V_0}$. However, it is intractable to search over $\scF_{V_0,N}$ as there are exponential amount of ways to choose $N$ nonzero parameters in $\scF_{V_0}$  

A reasonable approach is to first determine the non-zero parameter set. Unfortunately, from the proof of \cref{prop:gva}, one needs to find $F_0$ first, after which, matrix sparsification is applied. To achieve sparsification with linear order, the latter step relies on an intricate analysis of $L_{\tilde{H}}$ with the aid of barrier functions \cite{Lee15}. 

However, if we want to first apply matrix sparsification before solving the optimization problem, then a randomized edge selection heuristic is more suitable \cite{Spi11}. As a payoff, we can only hope to find $F_0 \in \scF_{V_0}$ with $\calO(|V_0|\ln|V_0|/\epsilon^2)$ parameters. For the rest of this section, we assume that $\scF_{V_0}$ is either choice~\ref{it:xtl} or \ref{it:xtlo}, and each matrix from choice~\ref{it:xym} is the difference of two matrices from choice~\ref{it:xtlo}. 

For randomized edge selection, we need the following notion.

\begin{Definition}
For a graph $G$ with Laplacian $L$, the \emph{effective resistance} $R_{u,v}$ between two vertices $u, v$ on $H$ is defined as: 
\begin{align*}
R_{u,v} = (\delta_u-\delta_v)'L^{+}(\delta_u-\delta_v), 
\end{align*}
where $\delta_u$ is the (column) vector taking value $1$ at $u$ and $0$ elsewhere, and $L^+$ is the pseudo-inverse of $L$ obtained by inverting non-zero eigenvalues of $L$.
\end{Definition}

Intuitively, effective resistance $R_{u,v}$ between two vertices $u,v$ gives a good measure on the connectivity between them, and hence we may use $R_{u,v}$ to determine whether their should be an edge between $u,v$. 

We briefly recall the steps for \emph{randomized edge sparsification} for a graph of size $m$:
\begin{enumerate}[i)]
\item For a pair of vertices $u,v$, compute the effective resistance $R_{u,v}$.
\item Let $w(u,v)$ be the edge weight between $u$ and $v$. Define 
\begin{align*}
p_{u,v} = \min\{1,4w(u,v)R_{u,v}\epsilon^{-2}\log m\}.
\end{align*}
\item An edge $(u,v)$ is included in the new sparse graph with probability $p_{u,v}$.
\end{enumerate}

To solve the problem in \cref{prop:gva}, we are required to perform (a) edge selection and (b) optimization simultaneously, causing exponential complexity. We notice that the above randomized edge selection only requires input from the graph. Therefore, we may alternate between edge selection and graph estimation. The main steps can be described as follows:
\begin{enumerate}[i)]
\item Set $G_0 = G$ and let $m = |V_0|$. Choose an integer parameter $N_1 = \calO(m\log m)$, whose exact size depends on error $\epsilon$. Let $N_2$ be a small fraction of $rN_1$, with $0<r<1$.
\item \label{it:agi} Assume $L_{G_i} \in \scF_{V_0}$ is determined for a graph $G_i$. Add small random edge weights to every pair $u,v \in V_0$ not connected by an edge in $H_0$ to obtain a graph $G_i'$.  
\item For $(u,v) \notin H_0$, compare the effective resistance $R_{u,v}$ in $G_i'$. Rank such pairs in descending order according to $w(u,v)R_{u,v}$, denoted by $Q$. 
\item \label{it:stt} Select the top $N_1$ pairs in $Q$ and randomly choose $N_2$ pairs from the remaining pairs in $Q$. The union of these $N_1+N_2$ pairs is denoted by $Q_i$.
\item Construct the subspace $\scF_{V_0, Q_i} \subset \scF_{V_0}$ parametrized by nonzero variables associated with pairs in $Q_i$.
\item $G_{i+1}$ is obtained by solving the optimization problem in \cref{prop:gva} with $F^*_0\in \scF_{V_0,Q_i}$.
\end{enumerate}

The steps are repeated for a fixed amount of iterations or until converge. It is worth mentioning that the purpose of Step~\ref{it:agi}-\ref{it:stt} is to ensure in each iteration, the nonzero parameter family $Q_i$ does not stuck in initial iterations.

\end{document}